\documentclass[twocolumn]{aa} 
\usepackage{graphicx}
\usepackage[varg]{txfonts}
\usepackage{natbib}
\usepackage[version-1-compatibility]{siunitx}
\newcommand{\norm}[1]{\left\lVert#1\right\rVert} 
\newcommand{\ramses}{{\tt RAMSES}}

\newcommand{\be}{\begin{equation}}
\newcommand{\ee}{\end{equation}}
\newcommand{\beal}{\begin{aligned}}
\newcommand{\eeal}{\end{aligned}}

\usepackage{hyperref}
\bibpunct{(}{)}{;}{a}{}{,} 
\usepackage{amsmath}
\usepackage{esdiff}
\usepackage{xcolor}
\usepackage[english]{babel}
\usepackage{multirow}

\usepackage{orcidlink}

\begin{document}

   \title{The role of magnetic fields in the formation of multiple massive stars}

   \author{R. Mignon-Risse\,\orcidlink{0000-0002-3072-1496}
          \inst{1,2}
          \and
          M. Gonz\'alez\,\orcidlink{0000-0002-3197-6095} \inst{2}
          \and
          B. Commer\c con\,\orcidlink{0000-0003-2407-1025} \inst{3}
          }

   \institute{Universit\'e Paris Cit\'e, CNRS,  AstroParticule et Cosmologie, F-75013, Paris \\
         \email{raphael.mignon-risse@apc.in2p3.fr}
         \and
         Universit\'e Paris Cit\'e, Universit\'e Paris-Saclay, CEA, CNRS, AIM, F-91190 Gif-sur-Yvette, France
         \and
         \'Ecole Normale Sup\'erieure de Lyon, CRAL, UMR CNRS 5574, Universit\'e Lyon I, 46 All\'ee d'Italie, 69364 Lyon Cedex 07, France
             }

   \date{Received ?; ?}
 
  \abstract
   {
   Most massive stars are located in multiple stellar systems. Magnetic fields are believed to be essential in the accretion and ejection processes around single massive protostars.
    }
   {Our aim is to unveil the influence of magnetic fields in the formation of multiple massive stars, in particular on the fragmentation modes and properties of the multiple protostellar system.
   }
   {Using {\tt RAMSES}, we follow the collapse of a massive pre-stellar core with (non-ideal) radiation-(magneto-)hydrodynamics.
   We choose a setup 
   which promotes multiple stellar system formation in order to investigate the influence of magnetic fields on the multiple system's properties.
    }
   {In the purely hydrodynamical models, we always obtain (at least) binary systems following the fragmentation of an axisymmetric density bump in a Toomre-unstable disk around the primary sink: this sets the frame for further studying stellar multiplicity.
   When more than two stars are present in these early phases, their gravitational interaction triggers mergers until there are two stars left only.
   The following gas accretion increases their orbital separation and hierarchical fragmentation occurs so that both stars host a comparable disk and stellar system which then form similar disks as well. 
   Disk-related fragmenting structures are qualitatively resolved when the finest resolution is ${\approx}1/20$ of the disk radius.
      We identify several modes of fragmentation: Toomre-unstable disk fragmentation, arm-arm collision and arm-filament collision. 
      Disks grow in size until they fragment and become truncated as the newly-formed companion gains mass.
      When including magnetic fields, the picture evolves: the primary disk is initially elongated into a bar, it produces less fragments, {disk formation and} arm-arm collision {are} captured at comparatively higher resolution, and arm-filament collision is absent.
      {Magnetic fields 
      reduce the initial orbital separation but do not affect its further evolution, which is mainly driven by gas accretion.}
      {With magnetic fields, the growth of individual disks is regulated even in the absence of fragmentation or truncation}.
   }
   {Hierarchical fragmentation is seen in unmagnetized and magnetized models. 
   Magnetic fields{, including non-ideal effects,} are important because they remove certain fragmentation modes and limit the growth of disks, which {is otherwise only limited through fragmentation.}
   }


   \maketitle
%

\section{Introduction}
\label{sec:intro}

Massive stars are very often located in multiple systems (\citealt{sana_binary_2012}, \citealt{chini_spectroscopic_2012}, \citealt{duchene_stellar_2013}), and the importance of magnetic fields for massive star formation has been growing recently (\citealt{girart_magnetic_2009}; for a review, we refer the reader to \citealt{tsukamoto_role_2022}), but the exact role of magnetic fields in the formation of multiple massive stars is not clear yet.
Disk fragmentation has been shown theoretically (\citealt{adams_eccentric_1989}, \citealt{shu_sling_1990}, \citealt{bonnell_massive_1994}) and now observed thanks to high angular resolution facilities such as ALMA (e.g. \citealt{ilee_g1192061_2018}, \citealt{johnston_spiral_2020}, \citealt{johnston_detailed_2020}).
In addition to their role in the origin of massive protostellar outflows predicted by \cite{kolligan_jets_2018}, \cite{commercon_discs_2022}, \cite{mignon-risse_collapse_2021-1}, \cite{oliva_modeling_2023} and detected with unprecedented resolution by \cite{moscadelli_snapshot_2022} using masers, magnetic fields are expected to limit gas fragmentation and therefore are certainly important for multiple massive star formation \citep{anez-lopez_role_2020}.
{Numerical studies had to tackle both core and disk fragmentation, since their relative contribution to the formation of multiple systems remains to be evaluated, both in the low-mass and high-mass stellar regimes.}
So far, {in the case of high-mass star formation, }the interplay between radiation on small scales and magnetic fields have been shown to limit core fragmentation (\citealt{commercon_collapse_2011}, \citealt{myers_fragmentation_2013}) in the ideal magneto-hydrodynamical (MHD) regime, i.e. with a perfect coupling between magnetic fields and the dust-gas-mixture.
{In the case of low-mass pre-stellar cores, this was shown in \cite{commercon_protostellar_2010}, following the study by \cite{hennebelle_magnetic_2008-1} without radiative transfer.
At the scale of turbulent molecular clouds, core fragmentation inhibition by magnetic pressure has been shown in a number of studies such as \cite{federrath_star_2012}; see also the review by \cite{padoan_star_2014} and more recently the study of \cite{lebreuilly_protoplanetary_2021}.} 
Incorporating ambipolar diffusion, a non-ideal MHD effect describing ion-neutral drift, the ability of magnetic fields to reduce disk fragmentation has been further shown in {the study of }\cite{mignon-risse_collapse_2021}{, oriented towards massive star formation}.

Observationally, the presence and influence of magnetic fields either in massive star formation or in multiple protostellar systems has been increasingly constrained thanks to polarization measurements.
The relative orientation of the field lines with respect to structures of interest (see e.g. \citealt{cox_twisted_2022}) has shown how the gas dynamics is coupled to the magnetic field topology.
Furthermore, in a hub-filament structure (a structure which could be crucial in general for massive star formation, see the review by \citealt{motte_high-mass_2018}), the magnetic fields strength has been found to be comparable to gravity (\citealt{wang_multiwavelength_2019}, \citealt{anez-lopez_role_2020}).
In the low-mass star formation context, the possible interplay between rotation and magnetic fields in gas fragmentation has been exposed by \cite{galametz_observational_2020} with the submillimeter array (SMA).
In addition, the magnetic field topology in a circumbinary disk around low-mass protostellar objects has also been revealed by \cite{alves_magnetic_2018} using ALMA polarization observations and shown to be active at launching outflows \citep{alves_molecular_2017}. It suggests a long-lasting influence over the entire protostellar phase.

The present paper builds on recent numerical results.
Disk fragmentation in a centrally-condensed system has been recently studied by \cite{oliva_modeling_2020} and \cite{mignon-risse_disk_2023}, hereafter Paper I.
Those have focused on the properties of gaseous fragments forming in an accretion disk around a massive protostar modelled as a sink cell or sink particle, respectively, for which spiral arms play a prominent role. 
However, those studies have explored one case in which a disk forms around a central massive protostar and neglected magnetic fields.
Hence, we propose to give a complementary point of view by focusing on the formation of multiple stellar systems (described by sink particles) and by studying the consequences of introducing magnetic fields.

Our choice of physical ingredients is well suited for studying protostellar disk fragmentation.
First of all, non-ideal MHD is found to impact the disk formation by circumventing the so-called magnetic braking catastrophe (see \citealt{wurster_role_2018} and references therein) and affecting subsequent disk properties that can impact disk fragmentation.
In fact, the fragmentation critical length (the so-called Jeans length, \citealt{jeans_vibrations_1902}) depends on local pressures: thermal pressure and magnetic pressure mainly.
It has been shown that ambipolar diffusion (drift between the ions and the neutrals) produces vertically thermally-supported disks rather than magnetically-supported as in the ideal MHD case (see e.g. \citealt{masson_ambipolar_2016} in the low-mass and \citealt{commercon_discs_2022} in the high-mass regime), of smaller size than in the hydrodynamical case.
Let us note that another non-ideal MHD effect, Ohmic dissipation produces a thermally-supported midplane, where fragmentation should occur, in the massive protostellar context \citep{oliva_modeling_2023-1}.
Moreover, including all three non-ideal MHD effects (ambipolar diffusion, Ohmic dissipation, Hall effect) in low-mass pre-stellar core collapse calculations, \cite{wurster_disc_2019} found less disk fragmentation in their magnetized models than in their purely hydrodynamical models.
For these reasons, we include MHD with ambipolar diffusion in our magnetized models, as those are already implemented in the \ramses{} code (\citealt{teyssier_cosmological_2002}, \citealt{masson_incorporating_2012}).
Finally, radiative transfer and its coupling to hydrodynamics are taken into account in order to accurately compute the gas temperature, essential for an accurate Jeans instability modelling.
In that view, we use \ramses{} with state-of-the-art physical ingredients: radiative transfer and MHD with ambipolar diffusion.

{The combination of these physical ingredients and astronomical unit-scale resolution complements other studies led so far, in particular those carried out on molecular cloud scales and on which stellar feedback by massive stellar clusters is crucial (e.g. \citealt{grudic_elephant_2018}, \citealt{ali_massive_2019}, \citealt{grudic_model_2020}).
Indeed, disk scales - a prerequisite for modelling disk fragmentation - are hardly reached in present-day molecular cloud scale simulations (the finest resolution is, for instance, $1000$~AU in \citealt{rosen_blowing_2021}, $100$~AU in \citealt{mathew_imf_2021}).
In \cite{bate_diversity_2018}, disk scales are reached but magnetic fields are neglected.
In \cite{lebreuilly_protoplanetary_2021}, disk scales are reached and magnetic fields with ambipolar diffusion are accounted for, but no massive star formation is reported (possibly due to limited statistics); moreover, the large (${>}100$) number of stars and the environmental turbulence significantly complicate the detailed analysis when it comes to disentangling  magnetic effects on the evolution of multiple stellar systems from other mechanisms, as mentioned above.
By focusing on the small scale physics, we can reveal potential dominant magnetic mechanisms in the formation and evolution of massive stellar systems; if identified, those should be modelled in future large-scale simulation studies for which the properties of stellar systems (mass, multiplicity, separation...) are important.}

{Finally, this setup's properties are also complementary to other studies of collapse of massive pre-stellar cores.}
While \cite{myers_fragmentation_2013} and \cite{mignon-risse_collapse_2021} have studied fragmentation in turbulent clouds, we propose here to consider a non-turbulent massive pre-stellar core with a rotation profile favoring fragmentation \citep{oliva_modeling_2020}.
The ability of rotation to promote fragmentation has been extensively shown in the literature (e.g. \citealt{machida_collapse_2005}, \citealt{wurster_disc_2019}).
Observational results suggested that core rotation could play a role in dragging the magnetic fields \citep{beuther_gravity_2020} and driving fragmentation even in turbulent cores (\citealt{girart_dr_2013}, \citealt{zhang_magnetic_2014}, with SMA).
Our consideration is also meant to explore other initial conditions that those studied previously, possibly co-existing in nature, but also making easier the identification of fragmentation modes and of the role played by magnetic fields than in turbulent simulations{, as mentioned above}.

The paper, which presents results obtained via numerical simulations of massive pre-stellar core collapse, is organized as follows.
First, we propose to study the influence of numerical resolution in Sec.~\ref{sec:resol}, on disk fragmentation and stellar multiplicity, to identify the scales of interest for fragmentation and the runs that are converged enough for a deeper physical analysis.
Then, based on these results, we study the influence of magnetic fields in Sec.~\ref{sec:mag}.
The next section presents the numerical methods. 
\\

\section{Methods}
\label{sec:model}

\subsection{Radiation-magnetohydrodynamical model}

Simulations are performed with the \ramses{} code (\citealt{teyssier_cosmological_2002}, \citealt{fromang_high_2006}).
\ramses{} is an adaptive-mesh refinement code (AMR) integrating the equations of radiation-MHD (RMHD).
Non-ideal MHD is taken into account in the form of ambipolar diffusion (ion-neutral drift, \citealt{masson_incorporating_2012}).
For the radiative transfert part, we use an hybrid radiative transfer method \citep{mignon-risse_new_2020}.
In this method, the M1 method (\citealt{levermore_relating_1984}, \citealt{rosdahl_ramses-rt:_2013}, \citealt{rosdahl_scheme_2015}) is employed to model the propagation and absorption of protostellar radiation, while the flux-limited diffusion (FLD, \citealt{levermore_flux-limited_1981}, \citealt{commercon_radiation_2011}, \citealt{commercon_fast_2014}) is used to model dust-gas emitted radiation.
The RMHD model consists in this set of equations
   \begin{equation}
   \begin{aligned}
   \diffp{\rho}{t} + \nabla \cdot [\rho \boldsymbol{u}] 
   &= 0, \\
   \diffp{\rho \boldsymbol{u}}{t} + \nabla \cdot [\rho \boldsymbol{u} \otimes \boldsymbol{u} + P \mathbb{I}]
   &= - \lambda \nabla E_\mathrm{fld} + \frac{\kappa_\mathrm{P,\star} \rho}{\mathrm{c}} \boldsymbol{F}_\mathrm{M1} +\boldsymbol{F}_\mathrm{L} \\
   & - \rho \nabla \phi, \\
   \diffp{E_\mathrm{T}}{t} + \nabla \cdot \biggl[\boldsymbol{u} \left( E_\mathrm{T} + P + B^2/2 \right) \\
   - (\boldsymbol{u} \cdot \boldsymbol{B}) \boldsymbol{B} - \boldsymbol{E}_\mathrm{AD} \times \boldsymbol{B} \biggr]
   &= - \mathbb{P}_\mathrm{fld} \nabla : \boldsymbol{u} + \kappa_\mathrm{P,\star} \, \rho \mathrm{c} E_\mathrm{M1}  \\
   &  \, \, \, \, \, \, - \lambda \boldsymbol{u} \nabla E_\mathrm{fld} + \nabla \cdot \left( \frac{\mathrm{c} \lambda}{\rho \kappa_{\mathrm{R,fld}}}\nabla E_\mathrm{fld} \right)\\
   &  \, \, \, \, \, \,  - \rho \boldsymbol{u} \cdot \nabla \phi, \\
   \diffp{E_{\mathrm{M1}}}{t}  + \nabla \cdot \boldsymbol{F}_\mathrm{M1}
   &= - \kappa_\mathrm{P,\star} \, \rho \mathrm{c} E_\mathrm{M1} + \dot{E}_\mathrm{M1}^\star, \\
   \diffp{\boldsymbol{F}_\mathrm{M1}}{t}  + \mathrm{c}^2 \nabla \cdot \mathbb{P}_\mathrm{M1}
   &= - \kappa_\mathrm{P,\star} \, \rho \mathrm{c} \boldsymbol{F}_\mathrm{M1}, \\
  \diffp{E_{\mathrm{fld}}}{t} 
  +  \nabla \cdot [\boldsymbol{u}  E_\mathrm{fld}]
  &=
    - \mathbb{P}_\mathrm{fld} \nabla : \boldsymbol{u} 
    + \nabla \cdot \left( \frac{\mathrm{c} \lambda}
   	{\rho \kappa_{\mathrm{R,fld}}} \nabla E_{\mathrm{fld}} \right)\\
    & \, \, \, \, \, \, +\kappa_{\mathrm{P,fld}} \, \rho \mathrm{c} \left( \mathrm{a_R} T^4 - E_{\mathrm{fld}} \right), \\
   \diffp{\boldsymbol{B}}{t} - \nabla \times \left[ \boldsymbol{u} \times \boldsymbol{B} + \boldsymbol{E}_\mathrm{AD} \right]
   &= 0, \\
   \nabla \cdot \boldsymbol{B} &= 0, \\
   \Delta \phi &= 4 \pi \mathrm{G} \rho,
   \end{aligned}
   \end{equation}
   where, $\rho$ is the gas density, $\boldsymbol{u}$ is the velocity vector, $P$ is the gas thermal pressure, $\lambda$ is the flux-limiter for the FLD module, $E_\mathrm{fld}$ is the FLD radiative energy, $\kappa_\mathrm{P,\star}$ is the Planck mean opacity computed at the effective temperature of the primary star, $\mathrm{c}$ is the speed of light, $\boldsymbol{F}_\mathrm{M1}$ is the M1 radiative flux, $\boldsymbol{F}_\mathrm{L} = (\nabla \times \boldsymbol{B}) \times \boldsymbol{B}$ is the Lorentz force, $\phi$ is the gravitational potential, $E_\mathrm{T}$ is the total energy which is defined as $E_\mathrm{T} = \rho \epsilon + 1/2 \rho u^2 + 1/2 B^2 + E_\mathrm{fld}$ (where $\epsilon$ is the specific internal energy), $E_\mathrm{M1}$ is the M1 radiative energy, $\boldsymbol{B}$ is the magnetic field, $\boldsymbol{E}_\mathrm{AD}$ is the ambipolar electromotive force, $\mathbb{P}_\mathrm{fld}$ is the FLD radiative pressure (which is not evolved as such but related to the other FLD moments, this is the FLD approximation), $\kappa_\mathrm{P,fld}$ is the Planck mean opacity in the FLD module (computed at the local gas temperature), $\kappa_\mathrm{R,fld}$ is the Rosseland mean opacity, $\mathrm{a_R}$ is the radiation constant, $\mathbb{P}_\mathrm{M1}$ is the M1 radiative pressure and $\dot{E}_\mathrm{M1}^\star$ is the injection term of the primary stellar radiation into the M1 module. For conciseness, we do not present the ambipolar electromotive force and the chemical network used to compute the resistivities \citep{marchand_chemical_2016}. For more details, we refer the reader to \cite{mignon-risse_collapse_2021}.

Coupling between the M1 and the FLD modules occurs through the term $\kappa_\mathrm{P,\star}\rho  \mathrm{c} E_\mathrm{M1}$ in  the equation of temporal evolution of the internal energy, which is given by
   \begin{equation}
    C_\mathrm{v} \diffp{T}{t} 
   = \kappa_\mathrm{P,\star} \, \rho \mathrm{c} E_\mathrm{M1}
   + \kappa_{\mathrm{P,fld}} \, \rho \mathrm{c} \left(E_{\mathrm{fld}} - \mathrm{a_R} T^4  \right).
   \end{equation}
   We use the ideal gas relation with the internal specific energy $\rho \epsilon = C_\mathrm{v} T$,
   where $C_\mathrm{v}$ denotes the heat capacity at constant volume.

\subsection{Initial conditions}
\label{sec:frag_ci}

We use similar initial conditions as Paper I and \cite{oliva_modeling_2020}, and refer the reader to these papers for a complete setup description.
We summarize its main characteristics here.
The massive pre-stellar core has a mass $M_\mathrm{c} = 200  \, \mathrm{M_\odot}$ and radius $R_\mathrm{c} = 20\,625 \, \mathrm{AU} = 0.1$~pc, with density profile 
$\rho(r) \varpropto r^{-3/2}$, resulting in a global free-fall time of $37$~kyr, but a free-fall time of a few kyr in the innermost regions.
Rotation is characterized by an angular frequency $\Omega (R) \varpropto R^{-3/4}$, where $R$ is the cylindrical radius, ensuring a rotational-to-gravitational energy ratio independent of the radius in the midplane of the core, equal to $5 \%$.
The initial temperature is $10$~K everywhere.
Hydrodynamical simulations are performed with the Lax-Friedrich solver and magnetohydrodynamical simulations are performed with the HLLD solver \citep{miyoshi_multi-state_2005}, as in \cite{mignon-risse_new_2020} and \cite{mignon-risse_collapse_2021}, respectively.

In the magnetohydrodynamical runs, a uniform vertical magnetic field is initialized (parallel to the rotation axis).
The mass-to-flux ratio divided by the critical mass-to-flux ratio \citep{mouschovias_note_1976} at the core border is $\mu=2$, corresponding to a magnetic field strength $\mathrm{B_0}=0.43mG$.
This corresponds to a strongly magnetized pre-stellar core, but since the magnetic field is uniform, the magnetization at the center of the cloud is not as strong as $\mu=2$ (see \citealt{commercon_discs_2022} and \citealt{mignon-risse_collapse_2021}).
Within a distance of ${\sim}200$~AU to the core center, the ratio of angular velocity to magnetic fields strength exceeds the critical value of $(\omega/B)_\mathrm{crit} = 3.19 \times 10^{-8} \mathrm{c_s}^{-1} \, \mathrm{yr}^{-1}  \mathrm{\mu G}^{-1}$ with $\mathrm{c_s}$ the sound speed given in $\mathrm{km \, s^{-1}}$ \citep{machida_collapse_2005-1}.

Following from Paper I, a sink with mass $0.01\, \mathrm{M_\odot}$ is initially present at the center of the cloud for all runs.
However, it is not fixed in space and other sink particles are allowed to form (see below).
Qualitatively-similar results are obtained without such a central sink initially because of the mechanisms presented in Sec.~\ref{sec:mult}. This aspect is discussed in Sec.~\ref{sec:comp_work}.

\subsection{Resolution and sink particles}
\label{sec:resol1}

The coarse resolution is level $5$ (equivalent to a $32^3$ regular grid) and we vary the finest resolution between levels $13$ and $16$, resulting in a physical resolution of $10$~AU to $1.25$~AU.
Eight runs are considered : four hydrodynamical runs (prefix {\tt HYDRO-}) and four magnetohydrodynamical runs (prefix {\tt MU2-}).
Suffix "LR" refers to "low resolution" ($10$~AU), "MR" to "mid-resolution" ($5$~AU), "HR" to "high resolution" ($2.5$~AU) and "VHR" to "very-high resolution" ($1.25$~AU).
There is no constraint on the sink position neither on the maximal number of sinks to form, unlike the fiducial run of Paper I.
As mentioned in Sec.~\ref{sec:frag_ci}, one sink is initially present at the center of the box.

{On all AMR levels below the maximum, c}ells are refined so that the Jeans length is resolved by $12$ cells (see \citealt{truelove_jeans_1997}).
At the finest level, sink particles \citep{bleuler_towards_2014} are introduced where a dense clump is found to be bound and Jeans unstable (more details in \citealt{commercon_discs_2022}). 
Sinks only interact gravitationally with the surrounding gas and other sinks, and merge if part of their radius (four cells here) overlap. 
Accretion onto sinks occurs if gas within the sink cells is above a density threshold given by $1.2 \times 10^{-13} (\mathrm{dx}/10\mathrm{AU})^{-15/8} \,  \mathrm{g\, cm^{-3}}$ (Eq.~11 of \cite{hennebelle_what_2020}, see also Paper I), where $\mathrm{dx}$ is the finest physical resolution of the simulation.

\begin{table}
\caption{Finest spatial resolution and magnetization level of the different runs. The cost of each run is given in Appendix~\ref{app:cost}.}
\label{table:ics}
\centering 
\begin{tabular}{c | c c } 
	\hline\hline
	 Model & $\Delta x$ [AU] &  $\mu$  \\ \hline 
	{\tt HYDRO-LR}   	&  $10$	 & 	 \multirow{4}{*}{$\infty$} \\ \cline{1-1} 
	{\tt HYDRO-MR}		 &  $5$   	 & 					   \\ \cline{1-1}
	{\tt HYDRO-HR}		 &  $2.5$   & 					   \\  \cline{1-1}
	{\tt HYDRO-VHR}	 &  $1.25$   & 					   \\  \hline
	{\tt MU2-LR}    		&   $10$   &     \multirow{4}{*}{$2$}	 \\ \cline{1-1}
	{\tt MU2-MR}   		 &  $5$    &    					   \\ \cline{1-1}
	{\tt MU2-HR}   		 &  $2.5$   &    					\\ \cline{1-1}
	{\tt MU2-VHR}   		 &  $1.25$   &    				  \\ \hline\hline
\end{tabular}
\end{table}

\section{Influence of spatial resolution}
\label{sec:resol}

In this section, we investigate the impact of spatial resolution on massive stellar multiplicity, i.e. the number of sink particles, and related properties. 

\subsection{On the presence of accretion structures}
\label{sec:resolstruc}

\begin{figure*}
\centering
    \includegraphics[width=18cm]{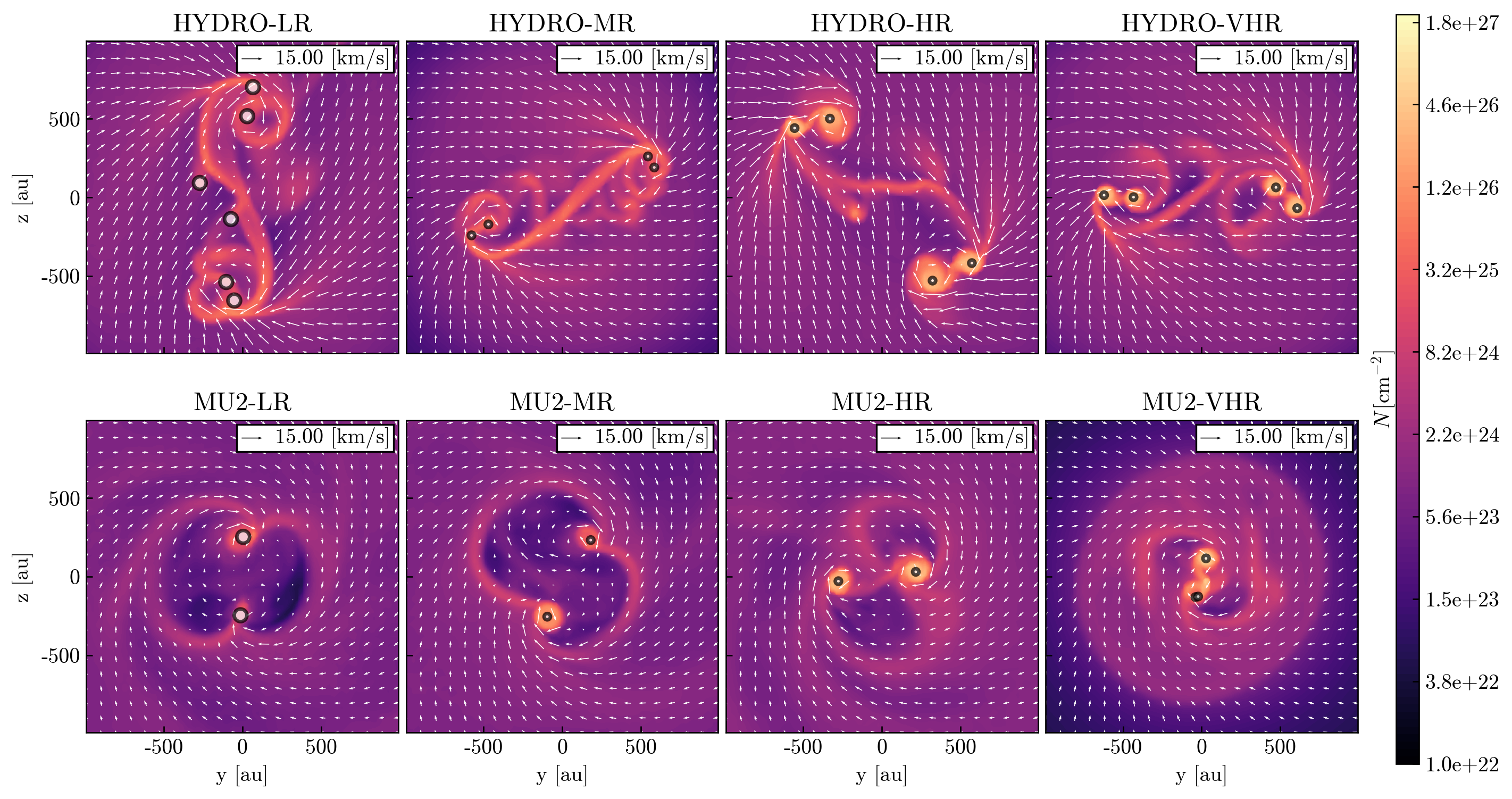}  
    \caption{Column density computed over $200$~AU along the $x-$axis (the rotation axis) in the hydro runs (top) and in the magnetic runs (bottom) at low resolution (left panels), mid-resolution (center-left), high resolution (center-right), and very high resolution (right panels), displayed in the $x=0$ plane, at the end of the runs. White dots indicate sink particles positions. Gas velocity vectors are overplotted.}
    \label{fig:coldens}
\end{figure*}

Figure~\ref{fig:coldens} shows the column density in the eight runs at the end of the run, ranging from $t=11$~kyr for run {\tt MU2-VHR}(about a quarter of the core free-fall time) to $t=20$~kyr for the hydrodynamical runs.
To begin with, in the hydro case we report spiral arms and disk-like structures in run {\tt HYDRO-LR}, with a filament linking two stellar systems.
However, some sink particles do not possess disk.
Going to higher resolution, runs {\tt HYDRO-MR}, {\tt HYDRO-HR} and {\tt HYDRO-VHR} suggest convergence with respect to those structures, and exhibit also the spiral arms after they grow. 
In those runs, all sink particles have a disk.
Nevertheless, the individual disks clearly visible in run {\tt HYDRO-HR} (and {\tt HYDRO-VHR}) appear in the place of circumbinary disks in run {\tt HYDRO-LR}.
We note that the filaments between the primary sink (already present in the initial conditions) and the secondary sink (i.e. the first sink that forms in the simulation, since the primary is already there) were created concomitantly with the secondary sink particle.

We aim to check whether a given spatial resolution allows to capture spiral arm formation. In the hydrodynamical case, a $10$~AU resolution is sufficient to capture structures prone to fragmentation, although circumstellar and circumbinary disks can be confused, and a $5$~AU to $1.25$~AU resolution appears to give a converged qualitative picture.

In all magnetic runs, a pseudo-disk is present.
It is characterized by an enhanced density compared with the surrounding medium, a magnetic pressure dominating over thermal pressure, and infall motions. 
Its size increases with time.
In Fig.~\ref{fig:coldens} it is clearly visible for run {\tt MU2-MR} and {\tt MU2-HR} with a ${\sim}1000$~AU radius and for run {\tt MU2-VHR} with a ${\sim}700$~AU radius.
We will not focus on pseudo-disks \citep{galli_collapse_1993} but rather on rotationally-supported disks in which the centrifugal acceleration ensures radial equilibrium (simply called "disks" for the rest of the paper).
As can be seen in the bottom-left panel of Fig.~\ref{fig:coldens}, in run {\tt MU2-LR} there is no rotating disk larger than the sink particle size at $t=10$~kyr.
Consequently, there is no spiral arm or filament connecting the two sinks or their surrounding accretion structures. In this run, disks appear after about half of the simulation time. 
In runs {\tt MU2-MR}, {\tt MU2-HR} and {\tt MU2-VHR}, we find individual disks, spiral arms and a filament linking the two disks, as in the hydrodynamical case. In run {\tt MU2-MR}, individual disks are resolved but over the integration time we do not observe the formation of new spiral arms after a binary has formed. On the opposite, we report that spiral arms keep forming regularly in runs {\tt MU2-HR} and {\tt MU2-VHR}, although they are hardly distinguishable on a single time snapshot like Fig.~\ref{fig:coldens}. 
In run {\tt MU2-VHR}, a third sink particle forms in one of the individual disks after spiral arm collision (reminiscent of \citealt{mignon-risse_collapse_2021}).
The common accretion disk, where the third sink formed and is still embedded at the end of the run, is the closest structure to what could be called a circumbinary disk in any of these simulations.
For the magnetic case, under these conditions we find that a spatial resolution of $2.5$~AU is sufficient to resolve the structures prone to fragmentation - which does not mean that convergence is reached, see Sections~\ref{sec:lj} and \ref{sec:mult}.
In particular, disks (more easily identifiable by their column density structure than by rotation arguments, see \citealt{mignon-risse_collapse_2021}) are smaller than in the hydrodynamical case (see Sec.~\ref{sec:diskrad}; and smaller than in \citealt{mignon-risse_collapse_2021}). 
Consequently, spiral arms and fragmentation induced by spiral arm collision appear at higher resolution than in hydrodynamical runs, and fragmentation might be missed in simulations that do not reach such spatial resolutions requirements.

Let us note that the cavities visible in the bottom-left panel of Fig.~\ref{fig:coldens} (run {\tt MU2-LR}, also visible later in the other magnetic runs) are reminiscent of those reported in \cite{hennebelle_what_2020}, \cite{commercon_discs_2022}, and \cite{mignon-risse_collapse_2021}. These are likely due to the interchange instability (see Appendix~B of \citealt{mignon-risse_collapse_2021}).

\subsection{Jeans length}
\label{sec:lj}

\begin{figure}
\centering
    \includegraphics[width=8cm]{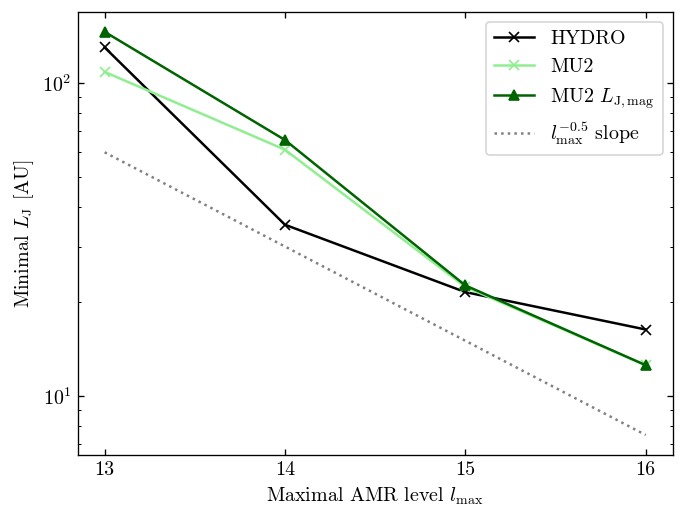}  
    \caption{Minimal Jeans length as a function of the maximal AMR level $l_\mathrm{max}$ (a higher AMR level means a finer resolution), computed at $t=10$~kyr: each point refers to a simulation of Table~\ref{table:ics}. In the magnetic case, we compute the thermal ($L_\mathrm{J}$, see Eq.~\ref{eq:nj}) and magnetic ($L_\mathrm{J,mag}$) Jeans length. An additional AMR level gives a finer resolution by a factor of two.}
    \label{fig:lj}
\end{figure}


After we have addressed the spatial resolution needed to capture structures of interest, let us see how resolution impacts the Jeans length.
The (thermal) Jeans length is defined as 
\be
L_\mathrm{J} = \frac{ \sqrt{\pi} c_\mathrm{s} }{\sqrt{\mathrm{G} \rho}},
\label{eq:nj}
\ee
where $c_\mathrm{s}= \sqrt{\gamma P/ \rho}$ is the local sound speed, with $\gamma=5/3$ (in our calculations) the adiabatic index. The magnetic Jeans length $L_\mathrm{J,mag}$ is defined in a similar fashion but using $\sqrt{ c_\mathrm{s}^2 + v_\mathrm{A}^2 }$, where $v_\mathrm{A}=B/\sqrt{4\pi \rho}$ is the Alfv\'en speed, instead of $c_\mathrm{s}$ (more details in Sec.~\ref{sec:mag} on the comparison between $L_\mathrm{J}$ and $L_\mathrm{J,mag}$).
For an alternative approach to computing the Jeans length in the MHD case, 
we refer the reader to the Appendix of \cite{myers_fragmentation_2013}, who also discuss the anisotropy of magnetic effects.
Note, however, that the difference between the $L_\mathrm{J,mag}$ they derive and our simple formulation is minor.

Figure~\ref{fig:lj} shows the minimal Jeans length as a function of the maximal AMR level (a higher AMR level means a finer resolution by a factor of two).
Hence, one would like the Jeans length slope to be less steep than a $l_\mathrm{max}^{-0.5}$ slope and ideally to become flat, indicating that the Jeans length becomes increasingly well sampled as the resolution increases.

It is visible in Fig.~\ref{fig:lj} that for certain AMR levels, the smallest $L_\mathrm{j}$ actually decreases faster than a $l_\mathrm{max}^{-0.5}$ slope.
It is particularly visible from $l_\mathrm{max}=13$ to $l_\mathrm{max}=14$ (runs {\tt HYDRO-LR} and {\tt HYDRO-MR}).
Indeed, we find that the maximal density reached increases with the resolution (not shown here for conciseness). Consequently, a {higher maximum AMR level} does not always result, in those simulations, in a better-sampled Jeans length.
Nevertheless, we checked that the minimal Jeans number, that is the number of cells sampling a Jeans length, is generally larger than $8$ {at the maximum level}, so that {artificial} fragmentation should be avoided {(see the discussion in \ref{sec:comp_work})}.
There is no clear trend in the magnetic runs for a Jeans length generally decreasing less steeply than $l_\mathrm{max}^{-0.5}$.
Nevertheless, it is the case from $l_\mathrm{max}=14$ to $l_\mathrm{max}=16$ for the hydro runs, corresponding to a resolution of $2.5~$AU (run {\tt HYDRO-HR}) to $1.25$~AU (run {\tt HYDRO-VHR}).

{In summary}, checking the Jeans length sampling at a given resolution only indicates that fragmentation{, if reported,} is not {artificial} (\citealt{truelove_jeans_1997},{ see also \citealt{federrath_new_2011} and the discussion in Sec.~\ref{sec:comp_work}}).
{It does not mean that the Jeans length converges towards a unique value because it depends on the density and temperature, which depend on the maximum resolution.}

\subsection{Multiplicity: number of sink particles}
\label{sec:mult}

\begin{figure*}
\centering
    \includegraphics[width=8.5cm]{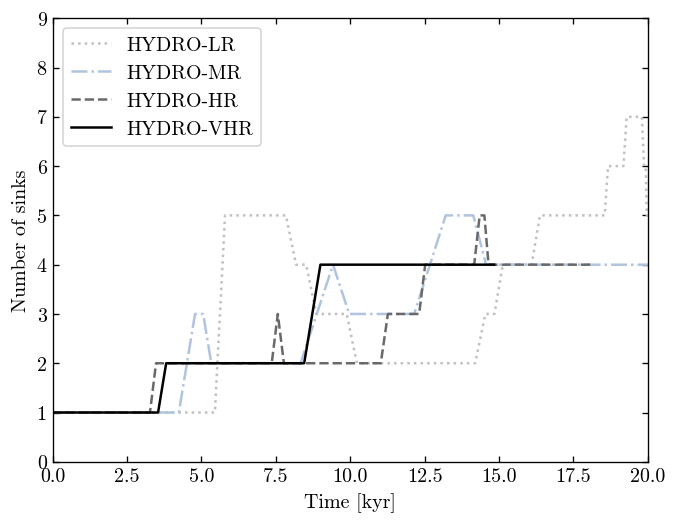}  
    \includegraphics[width=8.5cm]{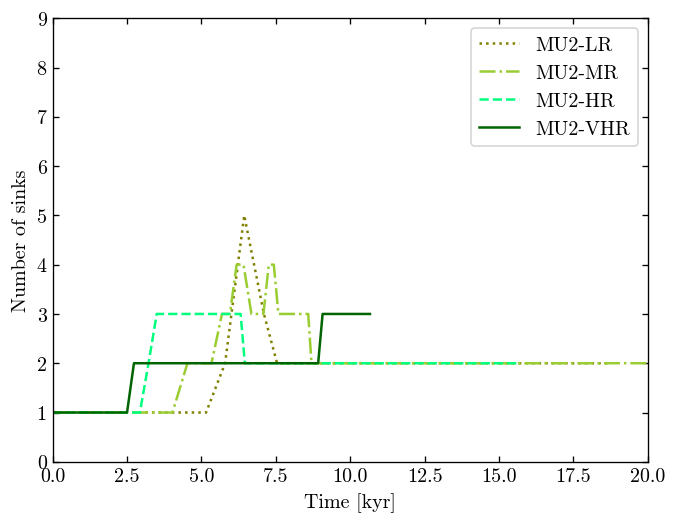} 
    \caption{Number of sink particles as a function of time in the hydro runs (left panel) and in the magnetic runs (right panel).}
    \label{fig:tnsink}
\end{figure*}

Figure~\ref{fig:tnsink} shows the number of sinks as a function of time. We distinguish three phases: initial fragmentation (increasing number of sinks) due to Toomre-unstable disk fragmentation, mergers (number of sinks decreasings towards a value of $2$), and secondary fragmentation.
In runs {\tt HYDRO-VHR} and {\tt MU2-VHR}, only two sinks are formed after the first fragmentation phase so there is no merger phase.

First, it can be seen that the sink multiplicity increases before $t = 6$~kyr from $1$ (the initial sink) to $2-5$. This is the initial fragmentation epoch, which is highly impacted by the formation of a disk with a density bump on top of it, prone to fragmentation as it is the most Toomre-unstable region (see Paper I). This structure fragments rapidly (in less than one orbital period, \citealt{norman_fragmentation_1978}) into four pieces in runs {\tt HYDRO-LR} and {\tt MU2-LR}, likely triggered by the Cartesian grid. 
{As already mentioned in Paper I, these grid effects dominate for smooth initial conditions; other possibilities would be to choose perturbed (e.g. \citealt{commercon_protostellar_2008}, \citealt{kuruwita_binary_2017}) or turbulent initial conditions (e.g. \citealt{gerrard_role_2019}, \citealt{kuruwita_role_2019}, \citealt{mignon-risse_collapse_2021}) to seed the instability.}
In the other runs, we nevertheless observe the development of non-axisymmetries similar to spiral arms around the primary sink, where additional sinks form.
In runs {\tt MU2-MR},  {\tt MU2-HR} and {\tt MU2-VHR}, the disk is affected by magnetic forces and is elongated into a bar, similarly to the structure reported in \cite{machida_collapse_2005} following \cite{machida_collapse_2005-1}.
Hence, in those cases, the disk instability has unlikely been seeded by the grid.

During a second phase, the sink multiplicity decreases as sinks merge. 
{The central sink was initially in an unstable equilibrium as it was surrounded by massive companions. Hence, any shift (e.g. caused by numerical errors or any asymmetry) leads this sink to leave the center.}
We understand the drop in sink multiplicity to be due to gravitational interactions between sinks.
Because of these interactions, the sinks' angular momentum (computed with respect to the origin) changes, favoring the migration of some sinks towards the center, where the primary sink is initially located, until they merge.
Eventually, all runs show a binary system at some point between $t=7.5$~kyr and $t=10$~kyr. 

\begin{figure}
\centering
    \includegraphics[width=8.5cm]{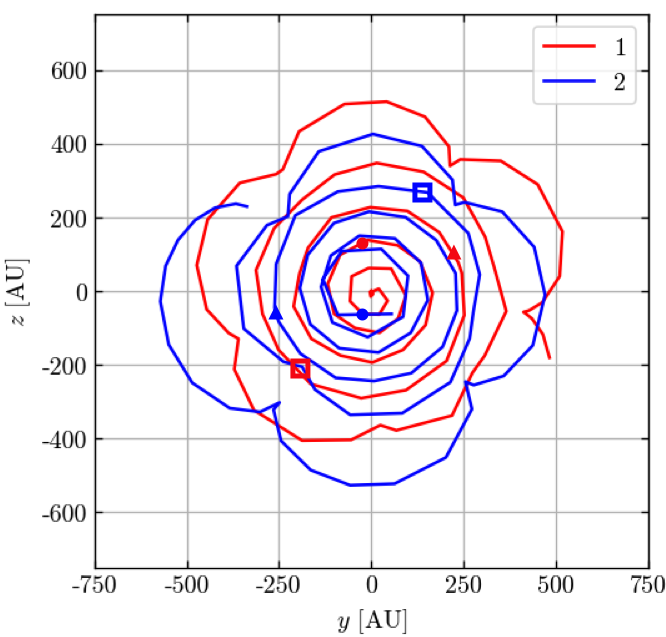}
    \caption{Trajectory of sinks $\#1$ and $\#2$ in run {\tt HYDRO-HR} in the $(y,z)$-plane (perpendicular to the initial rotation axis).
    The sinks location at $t=5$~kyr, $t=10$~kyr, and at the time of secondary fragmentation, is indicated by the dots, the triangles, and the empty squares, respectively.}
    \label{fig:t_yzsink1}
\end{figure}

In the hydrodynamical runs, the sink multiplicity increases again during the third phase, that we refer to as secondary fragmentation. At this stage, each sink of the binary system has developed an accretion disk. New sinks form at the extremity of spiral arms or following the collision between a spiral arm and a filament or another spiral arm. 
The final number of sinks is between $4$ (mid-, high- and very-high resolution) and $7$ (low-resolution). According to the value of the Jeans number being larger than $8$ in the {\tt HYDRO-LR} run and most fragmentation structures (circumbinary disk, filaments) being captured, fragmentation is physical rather than numerical. Nonetheless, the excess of sinks in run {\tt HYDRO-LR} can be explained by the lack of convergence regarding the resolution of structures.

In the {\tt MU2} runs, the number of sinks at the end of the simulations is between $2$ and $3$. In run  {\tt MU2-LR}, we have shown above that no clear disk structure forms around the two sinks (bottom-left panel of Fig.~\ref{fig:coldens}), hence we do not expect the formation of additional sinks. This system has reached a quiescent state. As shown in run {\tt MU2-VHR} (bottom-right panel of Fig.~\ref{fig:coldens}), secondary fragmentation is not fully prevented in the presence of magnetic fields, as we observe that a third sink particle has formed from spiral arms collision. It occurs in the close neighbourhood of an existing sink, at a distance of $19$~AU. This could not have occurred in run {\tt MU2-HR} with such a small orbital separation, because of the two sinks accretion radii ($10$~AU each).

\begin{figure}
\centering
    \includegraphics[width=8.5cm]{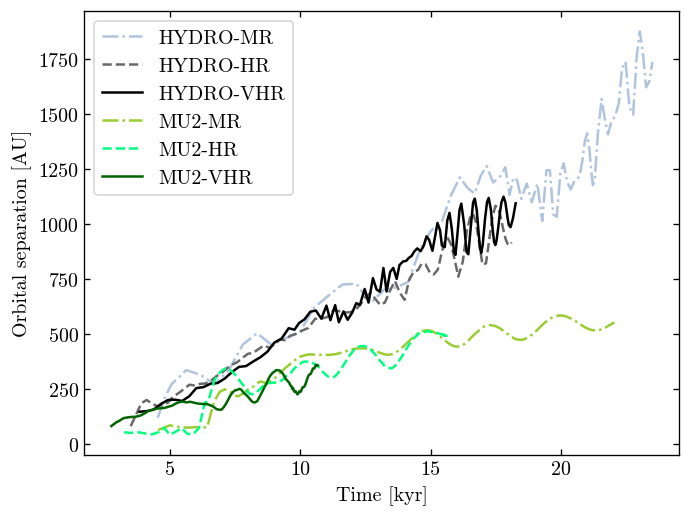}  
    \caption{Orbital separation between sinks $\#1$ and $\#2$ as a function of time.}
    \label{fig:t_b}
\end{figure}

Figure~\ref{fig:t_b} shows the orbital separation between the sinks that will eventually become the two most massive sinks as a function of time.
We find it to correctly converge with resolution.
It will be studied in more details in Sec.~\ref{sec:orbsep}.

Overall, we observe hierarchical fragmentation, with each sink possessing its own disk that can lead to additional fragmentation, in the hydrodynamical case as well as in the magnetic case.
We find that early fragmentation is compensated by enhanced stellar mergers following gravitational interactions and radial migration.
This suggests that high-multiplicity systems born from disk fragmentation likely formed out of hierarchical fragmentation, rather than from a single disk.

\subsection{Conversion of gas into stars}

\begin{figure*}
\centering
    \includegraphics[width=8.5cm]{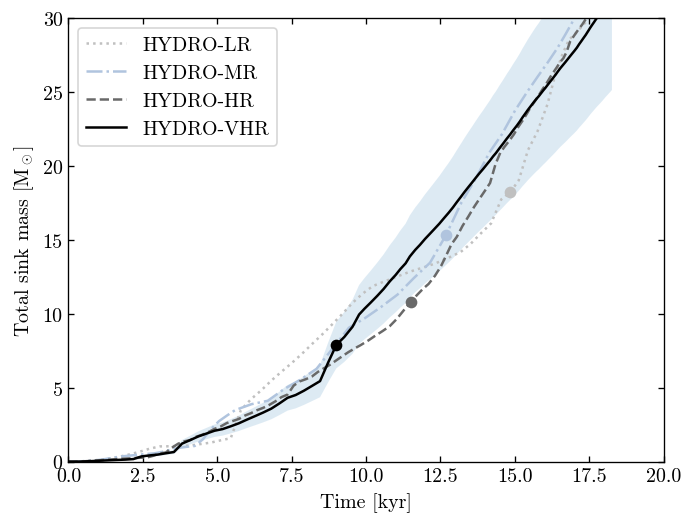}  
    \includegraphics[width=8.5cm]{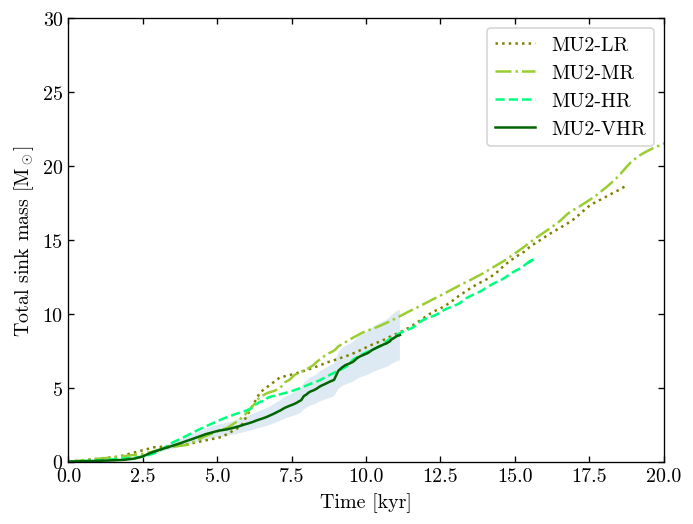}   
    \caption{Sum of all sink masses in the {\tt HYDRO} runs (left panel) and the {\tt MU2} runs (right panel) as a function of time. On the left panel, the filled circles indicate the secondary formation epoch.
    The blue region shows the value obtained in the highest resolution runs $\pm20\%$.}
    \label{fig:tmsink}
\end{figure*}

Figure~\ref{fig:tmsink} shows the sum of all sink masses as a function of time.
We observe a satisfying convergence in the {\tt HYDRO} run at the end of the simulation, within ${\approx}20\%$ with respect to the {\tt HYDRO-VHR} run. We nevertheless observe large variations at low resolution (run {\tt HYDRO-LR}).
Let us recall that the density threshold for accretion obeys a scaling relation \citep{hennebelle_what_2020}, which appears here to lead to self-consistent results.
In the magnetic runs, we find an agreemeent on the total mass within ${\approx}20\%$ as well at the end of the runs.

To sum up, the total mass is nearly independent of spatial resolution, especially at late times.

\section{Influence of magnetic fields}
\label{sec:mag}

In this section, we investigate the impact of magnetic fields on the fragmentation modes and on the properties of massive multiple stellar systems.

\subsection{Multiplicity and modes of fragmentation}
\label{sec:magmul}

To begin with, we observe two common fragmentation mechanisms between hydrodynamical and magnetic cases, namely Toomre-unstable disk fragmentation (with a transition through an elongated bar in high-resolution magnetic runs) and arm-arm collision.
Moreover, in all runs we observe sink mergers and the presence of a binary system after the early fragmentation epoch (Fig.~\ref{fig:tnsink}).
The difference is found in the secondary fragmentation phase, which does not exist in most of the magnetic runs, except for a sink particle born out of spiral arm collision in run {\tt MU2-VHR}, indicating that fragmentation is not totally suppressed by magnetic fields. 
Meanwhile, arm-filament collision, which is found to trigger fragmentation in the hydrodynamical runs, is not observed in any magnetic runs.
Indeed, filaments linking binaries (as also observed by, e.g. \citealt{sadavoy_dust_2018} in IRAS 16293-2422) are more diffuse when magnetic fields are included, as can be seen in Fig.~\ref{fig:coldens}, which may be due to additional magnetic pressure (as can be the case for larger-scale filaments, see e.g. \citealt{federrath_star_2012}, \citealt{gutierrez-vera_non-ideal_2022}) in those regions of smaller density than in the disks, where pressure support is mainly magnetic. By broadening the filaments, magnetic pressure prevents, indirectly, fragmentation induced by arm-filament collision.

As shown in Fig.~\ref{fig:lj}, fragmentation is not directly suppressed by magnetic pressure support - although, as discussed above, it prevents indirectly the fragmentation through arm-filament collision -, which has a minor impact on the Jeans length. 
Indeed, we find that regions of small Jeans length are as numerous in the hydrodynamical as in the magnetic cases. Similarly, within the magnetic runs, the minimal Jeans length and magnetic Jeans length are roughly equal because thermal pressure largely dominates over magnetic pressure. 
$L_\mathrm{J}$ and $L_\mathrm{J,mag}$ converge towards the same value at high resolution, while in run {\tt MU2-LR} no thermally-supported disks were formed at that time.
When looking at the Jeans length more generally, we find that $L_\mathrm{J}$ and $L_\mathrm{J,mag}$ differ where the Jeans length is already large and therefore where the gas is already stable against fragmentation. Hence, magnetic pressure is not responsible for these differences in sink multiplicity.
More than that, if the scales on which ambipolar diffusion is efficient are not sufficiently resolved, magnetic field piles-up which leads to overestimating the stabilizing role of magnetic pressure.

Overall, including magnetic fields conserve two modes of fragmentation: Toomre-unstable disk fragmentation and fragmentation associated to arm-arm collision.
However, no fragmentation induced by arm-filament collision is reported.

\subsection{On the variations of the orbital separation in binaries}
\label{sec:orbsep}

As shown in Fig~\ref{fig:t_b}, the orbital separation between the stars that will become the most massive ones, $a(t)$, is found to increase nearly linearly in all runs on timescales of kyr.
Quasi-periodic variations on timescales of the order of a kyr are linked to eccentricity (as in, e.g., \citealt{park_population_2022}).
Variations on shorter timescales, for example between $15$~kyr and $18$~kyr in run {\tt HYDRO-VHR}, are due to orbital motions with other companions, born from secondary fragmentation.
 In the following we study, on the one hand, the impact of magnetic fields on the orbital separation and, on the other hand, the two possible origins of this outspiral motion: gravitational torques from the gas and gas accretion.

Let us first present the equation describing the orbiral separation evolution of the binary.
Under the approximation of equal-mass binaries (which is the case most of the time in the hydro runs when two stars are present, and in the magnetic runs as well) and centrifugal equilbrium we have $a =16 j^2/(\mathrm{G} M)$, where $j$ is the specific angular momentum of the binary - also equal to that of each component of the binary - and $M$ is the total mass of the binary. Then, the temporal variation of the orbital separation can be derived as
\be
\diffp{a}{t}  = \frac{16}{\mathrm{G}} \left( \frac{2 j }{M} \diffp{j}{t}  - \frac{j^2}{M^2} \diffp{M}{t}  \right),
\ee
so that the condition for an increasing orbital separation is (the formulation is similar to, e.g. \citealt{tiede_gas-driven_2020})
\be
\frac{1}{j} \diffp{j}{t} > \frac{1}{2 M} \diffp{M}{t}.
\label{eq:orbsep}
\ee
When computing the evolution of the sinks angular momentum and mass over timescales of kyr, on which we observe an increasing separation (Fig~\ref{fig:t_b}), we find this condition to be met, which is physically consistent with the outspiral we observe.

\subsubsection{Impact of magnetic fields on the separation}

\begin{figure}
\centering
    \includegraphics[width=8.5cm]{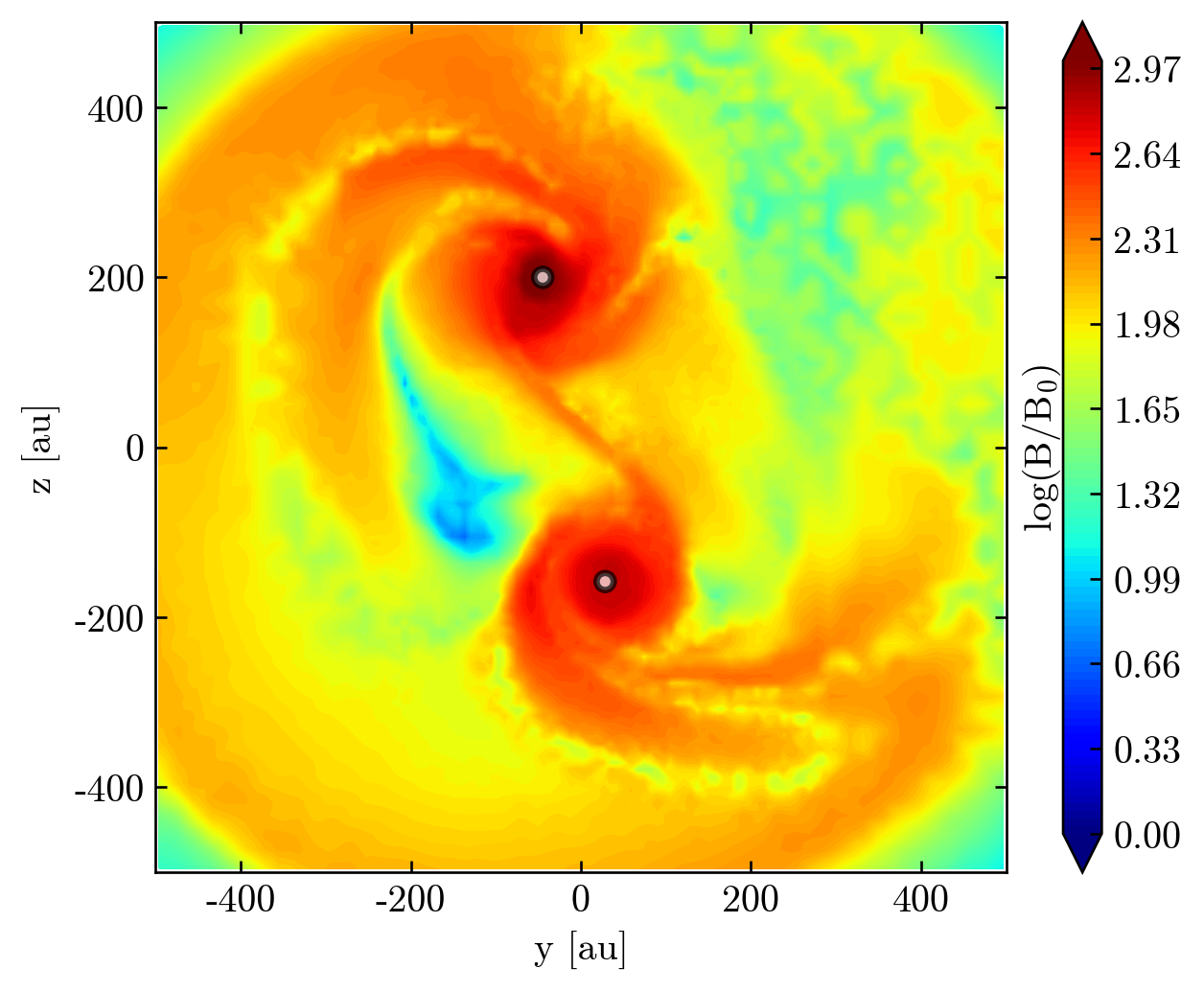}  
    \includegraphics[width=8.5cm]{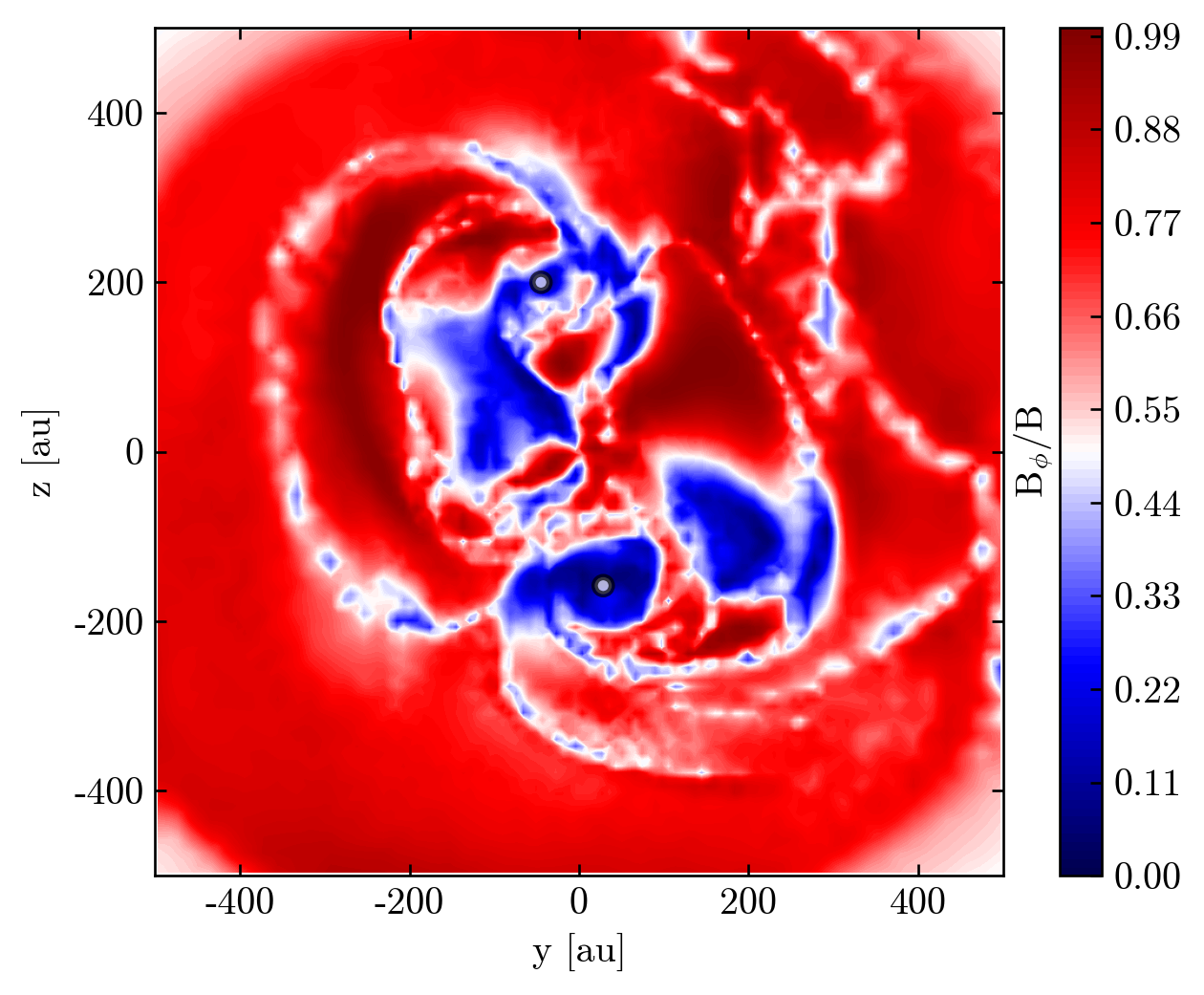}    
    \caption{Map of $B/\mathrm{B_0}$, where $\mathrm{B_0}$ is the initial, uniform magnetic field strength, and $B_\phi/B$ in the orbital plane in run {\tt MU2-HR} at $t=10$~kyr. Sink particles are denoted by white circles.}
    \label{fig:Bphi_B0}
\end{figure}

Do magnetic fields have any impact on the increasing separation? 
We observe that the orbital separation is reduced by a factor of two in binaries when magnetic fields are present.
Since $a(t)$ is nearly linear, it takes the form $a(t)=\alpha t$ where $\alpha$ is a constant, with $\alpha_\mathrm{HYDRO} \approx 2 \alpha_\mathrm{MU2}$.
Hence, the quantity $\dot{a}/a$ (where the dot indicates the time derivative) is equal in both cases.
{In other words, the orbital separation evolution is unaffected by magnetic fields.}

Magnetic fields only interact with sinks via their effect on the gas.
To address the possible angular momentum removal by magnetic braking, we compute the angular momentum at two distinct epochs, on cubes encompassing the binary system.

On the one hand, we focus on early times and small scales around the center of the cloud.
Computing the angular momentum in a cube of side $800$~AU centered onto the grid origin at $3.5$~kyr (typically the time when the secondary sink forms), we find that the specific angular momentum is reduced by $30\%$ in the magnetic case.
Hence, by the time the secondary sink forms, part of the angular momentum has been removed by magnetic braking.
This impact of magnetic fields explains the difference in the initial orbital separation of binaries.

On the other hand, we focus on later times, which means larger scales because the orbital separation increases.
Computing it in a cube of side $2000$~AU centered onto the grid origin at $t=5$~kyr and $10$~kyr, we do not find significant differences between the hydrodynamical and the magnetic runs.
How to explain the lack of efficiency of the magnetic braking (with respect to the center of mass, which is the center of rotation of the system)?
This mechanism's strength is proportional to the toroidal and poloidal components of the magnetic field, the former developing from coherent (differential) rotation.
Between $t=5$~kyr and $10$~kyr each star has a disk but no circumbinary disk. 
The consequence is twofold. 
First, as illustrated in the top panel of Fig.~\ref{fig:Bphi_B0}, the magnetic field strength outside the individual disks is smaller (by a factor of a few to a factor of ${\sim}100$) because the density there is smaller.
Second, portions of the gas rotating coherently around the binary indeed generate a toroidal field; locally, $B_\phi/\norm{\mathbf{B}}{\approx}1$, as shown in the bottom panel of Fig.~\ref{fig:Bphi_B0}.
However, these portions are too small, and the local magnetic field strength is too small for the overall magnetic braking to slow-down the rotating gas on timescales shorter than the time it takes for this same gas to reach any of the two individual disks.
In those disks, the magnetic braking with respect to the center of mass is negligible, as expected.
Hence, the binary system's angular momentum cannot be transported efficiently by magnetic fields.

Therefore, since $\dot{a}/a$ is equal in equal in both cases, this suggests that the value of $a(t)$ is inherited from its initial value, which differs in the hydrodynamical and in the magnetic cases because of magnetic braking.
{Hence, magnetic fields reduce the initial separation but not its further evolution.}

\subsubsection{Gravitational torques contribution}

\begin{figure}
\centering
    \includegraphics[width=8.5cm]{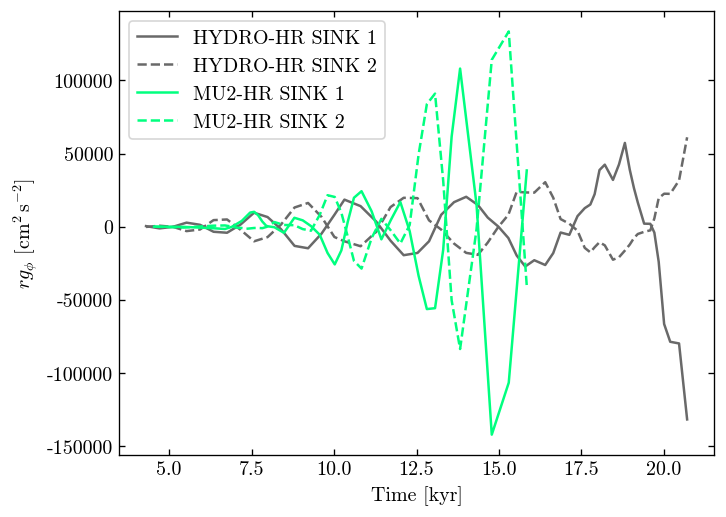}  
    \caption{Gravitational torque $r g_\phi$ acting on binaries in runs {\tt HYDRO-HR} and {\tt MU2-HR} as a function of time.}
    \label{fig:t_gphi}
\end{figure}

Now, one should investigate the contribution from gravitational torques and gas accretion onto this increasing separation.
To probe the former, we compute the sink specific angular momentum $j$ and the gravitational torques from the gas onto the sinks, noted $r g_\phi$, which accounts for the Plummer softening specific to gas-sink interactions.
The specific angular momentum globally increases with time - with values of the order of $10^{18} \mathrm{cm^2/s}$, and slope of ${\gtrsim}10^6  \mathrm{cm^2/s^2}$.
This is expected because the orbital separation increases with time whereas the binary mass, which can only increase, promotes orbital shrinking.
Theoretically, the gravitational torque must be negative when corresponding to the disk part lagging behind a sink and positive when it is in front of it (see e.g. \citealt{tiede_gas-driven_2020}).
If the gas distribution was perfectly symmetric with respect to the binary axis, the sum of the two should be zero. 
Meanwhile, spiral arms, companions (when present) and low-density cavities due to the interchange instability (in the magnetic models) should make it deviate from zero.
This is supported by our results: the total gravitational torque $r g_\phi$, shown in Fig.~\ref{fig:t_gphi} for runs {\tt HYDRO-HR} and {\tt MU2-HR}, starts showing an oscillatory-like behaviour, at about the orbital period, when spiral arms develop and a companion forms (in the hydro runs), as those orbit around the binary and alternatively contribute to a positive and negative torque.
On the opposite, low-density cavities always lag behind the sinks; their presence should translate into a positive torque on the sinks; since we find the torque to oscillate around zero, this contribution appears to be negligible compared to the spiral arm contribution.
We note, however, that the oscillation amplitude is larger in the magnetic runs than in the hydro runs.
This is likely due to the disks being smaller in the magnetic case: the distance to the entire spiral arm structure is smaller and the associated gravitational torque, which decreases quadratically with the distance, is larger.
Overall, the envelope of the oscillations in $r g_\phi$ has an amplitude of ${\approx}10^5 \mathrm{cm^2/s^2}$.
This is one order of magnitude too small to explain the slope of the specific angular momentum evolution.
Hence, only gas accretion can explain the increasing orbital separation.

\subsubsection{Gas accretion}

As shown in Eq.~\ref{eq:orbsep}, gas accretion can, depending on its specific angular momentum and mass, both contribute to increasing or decreasing the separation.
However, in this run the general trend is an increasing separation.
It is therefore primordial to understand why and to see whether it is purely linked to our setup.

Integrating both sides of Eq.~\ref{eq:orbsep} we get $\mathrm{d}j/j > \mathrm{d}M/2 M$.
Then, we integrate the left-hand side from $j_\mathrm{0}$ to $j'$ and the right-hand side from $M_\mathrm{0}$ to $M'$, where $j_\mathrm{0}$ (resp. $j'$ is the sink specific angular momentum prior to (resp. after) accretion and $M_\mathrm{0}$ (resp. $M'$) is the sink mass prior to (resp. after) accretion, yielding
$j'/j_\mathrm{0}> \sqrt{M'/M_\mathrm{0}}$.
To go further, we express $j'$ and $M'$ as functions of $j_\mathrm{0}$, $M_\mathrm{0}$ and the accreted gas angular momentum and mass labelled $j_\mathrm{gas}$ and $M_\mathrm{gas}$, as
\be
j' = j_\mathrm{0} \frac{
1+ \frac{j_\mathrm{gas}M_\mathrm{gas}}{j_\mathrm{0}M_\mathrm{0}}}
{1+ \frac{M_\mathrm{gas}}{M_\mathrm{0}}
}
\ee
and $M' = M_\mathrm{gas}+M_\mathrm{0}$, that we replace in the previous inequality.
After some simple algebra, as long as the gas mass increment is negligible compared to the sink mass ($M_\mathrm{gas} {\ll} M_\mathrm{0}$) so we neglect second-order terms $(M_\mathrm{gas}/M_\mathrm{0})^2$, this leads to the following criterion for an increasing orbital separation:
\be
j_\mathrm{gas} > \frac{3}{2} j_\mathrm{0},
\ee
where $j_\mathrm{0}=\sqrt{a \mathrm{G} M}$ from radial equilibrium between the sinks.
This is a reformulation of the trend found in the pioneer studies of \cite{bate_accretion_1997}, \cite{bate_predicting_2000}.
Putting in typical values of separation and binary mass from the simulations, we find that gas accretion increases the separation of the binary if
\be
j_\mathrm{gas} \, {\gtrsim} \, 2 \times 10^{21} \left( \frac{a}{200 \mathrm{AU}} \right)^{1/2} \left( \frac{M}{5\mathrm{M_\odot}} \right)^{1/2} \, \mathrm{cm^2 \, s^{-1}}.
\label{eq:jgas}
\ee

We will use those fiducial values for the following reasoning.
A relevant question is the following: does the gas, initially located at a given radius and with a given angular momentum (somewhat arbitrary in those simulations), possess enough angular momentum to fulfill condition \ref{eq:jgas} and therefore increase the orbital separation of the binary, at the time of interest?
For our initial rotation profile, the fiducial value of Eq.~\ref{eq:jgas} corresponds to gas initially located at radii $>10^3$~AU.
The follow-up question, to ensure the validity of the reasoning, is: does the gas initially located at this radius had enough time to reach the binary ? Yes, based on its associated free-fall time.
The scenario of an orbital separation driven by gas accretion in our simulations is thus consistent.

The next natural question would be how this value depends on the initial profile.
Although this comparison should be taken with caution because a different rotation profile would likely result in a distinct evolution of the system (e.g. \citealt{bate_predicting_2000}), for different initial rotation profiles such as the {\tt slow} and {\tt fast} models presented in \cite{commercon_discs_2022} the gas should be initially located at ${\sim}10^4$~AU to drive the outspiral of such a binary.
\newline

To sum up on this subsection dedicated to the orbital separation evolution, we find that gas accretion is mainly responsible for the binary outspiral and we can link such a trend to the initial rotation profile of the core.
We report the influence of magnetic fields in only one aspect: they remove the angular momentum in the innermost regions of the cloud at early times, setting a smaller initial orbital separation for binaries than in the hydrodynamical case.
However, they play no significant role in the following evolution of the orbital separation.

\subsection{Conversion of gas into stars}
\label{sec:B_conv}

As visible in Fig.~\ref{fig:tmsink}, in the hydrodynamical case, the total mass of all sinks is ${\sim}30\, \mathrm{M_\odot}$ near the end of the simulation ($t=17.5$~kyr) against ${\sim}15\, \mathrm{M_\odot}$ in the magnetic case. Nonetheless, the mass growth is similar within $30\%$ up to ${\sim}10$~kyr.

In the magnetic case the mass increases with a mean accretion rate slightly smaller than $10^{-3} \, \mathrm{M_\odot} \, \mathrm{yr^{-1}}$.
In the hydrodynamical case, the mean accretion rate is $\dot{M} {\approx} 10^{-3} \, \mathrm{M_\odot} \, \mathrm{yr^{-1}}$ up to the secondary fragmentation epoch (indicated by the circles in Fig.~\ref{fig:tmsink}). 
After that, it is $\dot{M} {\approx} 3.5  \times 10^{-3} \, \mathrm{M_\odot} \, \mathrm{yr^{-1}}$.
Hence, the change in the mean accretion rate occurs just before secondary fragmentation occurs.
This is consistent with the fragmentation being linked to spiral arms. Indeed, they carry angular momentum outwards, allowing the central object to accrete.
Though, it does not explain why the accretion rate remains at this higher value.
A possibility would be the initial density profile $\rho(r) \varpropto r^{-1.5}$ making more gas available for accretion at later times.
However, this would not explain the difference between the hydrodynamical and the magnetic case.
Moreover, this trend is not observed in the run with a single sink of Paper I (neither in the study of \cite{oliva_modeling_2020} with a similar initial density profile).
Hence, we do not attribute the enhanced accretion to the initial density profile.
Another possibility is that the presence of new close (${\sim}100$~AU) companions also triggers the formation of spiral arms on shorter timescales (the orbital timescale between the two close components), encouraging the accretion.
Similarly, the sinks born during secondary fragmentation form their own disk with its own spiral arms. A disk and a sink can perturb the secondary disk either via their spiral arms and via tidal forces (see e.g. \citealt{savonije_tidally_1994}), respectively. As a consequence, the other disk grows spiral arms and accretion would be enhanced. 

\begin{figure*}
\centering
    \includegraphics[width=8.5cm]{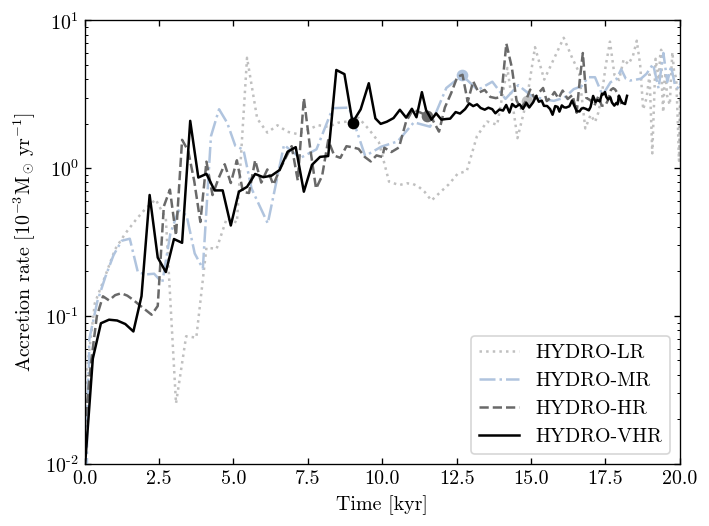}  
    \includegraphics[width=8.5cm]{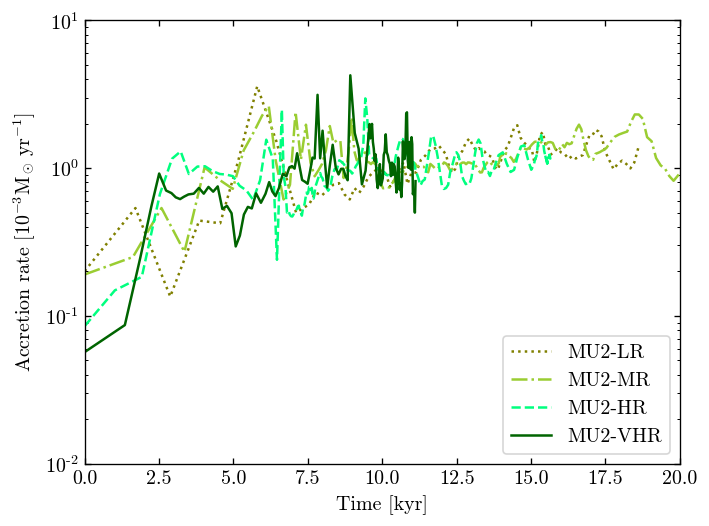}   
    \caption{Accretion rate in the {\tt HYDRO} runs (left panel) and the {\tt MU2} runs (right panel) as a function of time. On the left panel, the filled circles indicate the secondary formation epoch.}
    \label{fig:tmdot}
\end{figure*}

\begin{figure}
\centering
    \includegraphics[width=8.5cm]{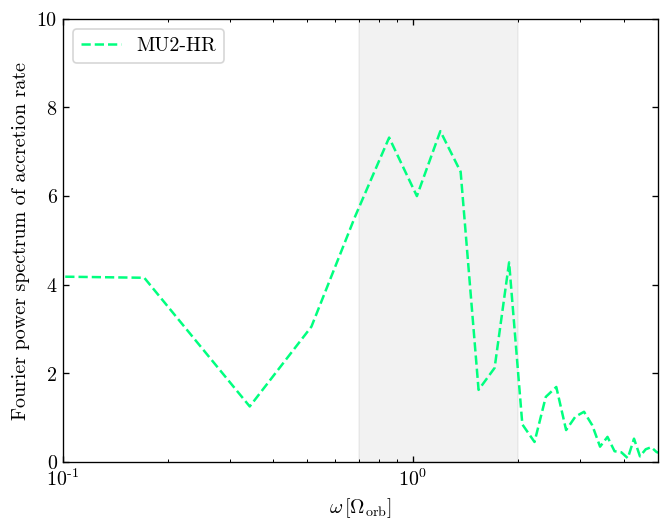}  
    \caption{Fourier power spectrum of the accretion rate in run {\tt MU2-HR} after $10$~kyr. The $x-$axis is the Fourier transform frequency normalized by the approximated orbital frequency $\Omega=1/2.5 \, \mathrm{kyr^{-1}}$. Data (instantaneous accretion rates) have been linearly interpolated in order to have a uniformly-spaced sample to perform the fast Fourier transform. The gray band indicates the $0.7-2\Omega_\mathrm{orb}$ interval. 
    The low-frequency signal does not point toward a physical periodicity but to the sample time which is only $6$~kyr long. More interestingly, we find no significant peak at frequencies higher than $2\Omega_\mathrm{orb}$. 
    This suggests that the accretion rate is modulated by frequencies of the order of the orbital frequency $\Omega_\mathrm{orb}$.}
    \label{fig:fft}
\end{figure}

In the following, we investigate the role of sink multiplicity on the accretion rate using this figure and look for patterns.
Figure~\ref{fig:tmdot} shows the instantaneous accretion rate as a function of time.
In the magnetic case, and in particular for run {\tt MU2-HR}, we observe accretion peak-like features after $10$~kyr that could be periodic.
We extract their exact time by selecting the time of accretion rates that are local maxima and that are greater than the mean accretion rate in the simulation.
We find that the time between two maxima is between $0.75$~kyr and $1.0$~kyr.
This interval matches the semi-orbital period that can be extracted from the sink motions (also visible in the orbital separation plot, Fig.~\ref{fig:t_b}, thanks to the non-zero eccentricity of the system).
This is consistent with the $m=2$ mode expected to develop when the binaries are not too close with the spiral wave pattern speed being the orbital frequency \citep{savonije_tidally_1994}.
In order to check for this periodicity, we have computed the Fourier power spectrum of the accretion rate after $10$~kyr, as shown in Figure~\ref{fig:fft}. We find an enhanced signal at frequencies between $0.7\Omega_\mathrm{orb}$ and $2\Omega_\mathrm{orb}$. 
In particular, we find no significant periodicity above $2\Omega_\mathrm{orb}$.
Nonetheless, it is difficult to draw firm conclusions from this plot, even though it suggests a plausible periodicity in the signal. Indeed, we know that the system's orbital frequency is not constant with time, there may be a non-periodic part of the signal that is not strictly constant either (see Fig.~\ref{fig:tmdot}) and only a few (${\approx}3$) orbital periods are captured after $10$~kyr within the simulation time.
This shows, however, than a link between multiplicity and accretion rate is plausible, in addition to being justified theoretically.

In the hydrodynamical runs, we find that the timescale between such accretion peaks mainly lies in the $0.1-0.4$~kyr interval.
This is also consistent with the previous scenario, since a larger ($>2$) sink multiplicity involves more than one characteristic frequency, and in those runs there are also close binary systems with higher orbital frequencies than in the magnetic runs. Because there are several frequencies involved, the periodicity in the accretion rate are more difficult to extract and much less visible (Fig.~\ref{fig:tmdot}).

Our simulations do not show a direct influence of magnetic effects on the accretion rate.
However, they suggest a possible modulation of the accretion rate at the orbital period due to the gravitational influence of a companion. 
Nevertheless, the system on hand - more than two components, a variable orbital period with time, a binary embedded in a collapsing core - is not well suited to extract such quasi-periodic features.

\subsection{Individual disk radius}
\label{sec:diskrad}

In this Section, our aim is to estimate the radius of the individual disks surrounding sink particles and the underlying process regulating its size, emphasizing the differences observed between hydrodynamical and magnetic models.
Several diagnostics can be used, in principle, to derive the radius of a disk.

First, we consider using kinematic signatures, as is done observationally with velocity-positions diagrams (e.g.
\citealt{murillo_keplerian_2013}, 
\citealt{tobin_vlaalma_2020}).
One way to do it in the simulation is to compute the deviation with respect to Keplerianity (as done in Paper I) on a cell-by-cell basis, and to reduce this information to a single scalar by computing the mean or median azimuthal value and looking for a transition radius.
However, we find the Keplerianity to vary significantly between the cells as it is greatly affected by the stellar companions.
The same limitation is encountered when using the disk criteria presented in \cite{joos_protostellar_2012} and used, e.g., in \cite{mignon-risse_new_2020}.
Therefore, we choose not to focus on a kinematic definition of the disk, although our results still point to structures dominated by rotation.

\begin{figure}
\centering
    \includegraphics[width=8.5cm]{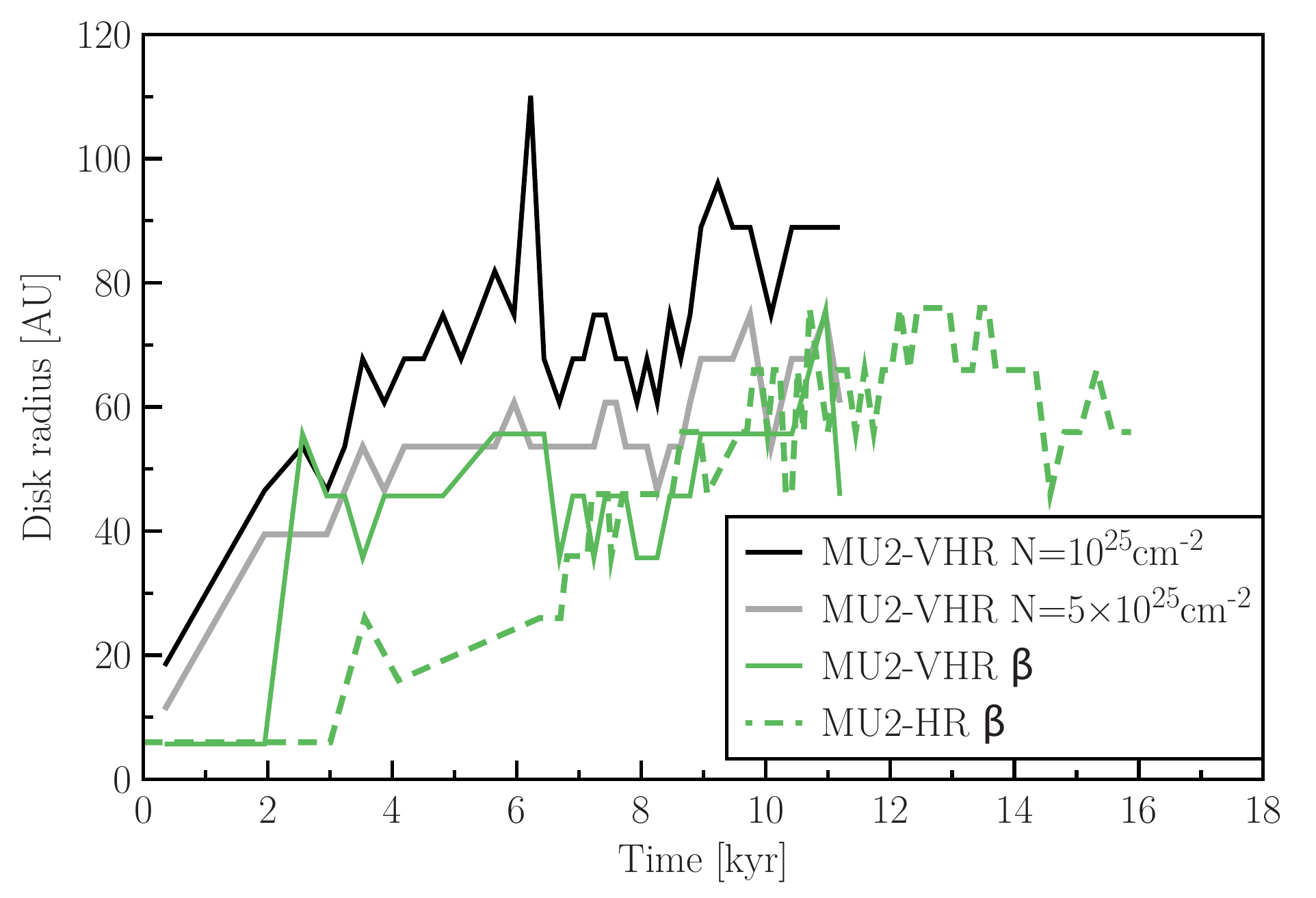}  
    \caption{Disk radius as a function of time, computed from the surface density with a threshold value $N=10^{25} \, \mathrm{cm^{-2}}$ (black curve), $N=5\times 10^{25} \, \mathrm{cm^{-2}}$ (gray curve) and the plasma beta (green curve) in run {\tt MU2-VHR}, and with the plasma beta in run {\tt MU2-HR} (green dashed curve).}
    \label{fig:rdisk}
\end{figure}

Observationally, dust continuum fitting is used as an alternative to kinematic signatures of disks (e.g. \citealt{patel_disk_2005}).
As in \cite{mignon-risse_collapse_2021}, we propose to use the surface density as a criterion to define the disk and thus, the disk radius.
Integrating the surface density over of depth of $200$~AU perpendicular to the disk plane (this is already sufficiently larger than the typical disk height, which is less than $50$~AU), we get a surface density $\Sigma(y,z)$, a two-dimensional function, as it depends on the spatial coordinates in the disk plane.
The largest surface density is found to be around $\Sigma {\sim} 10^{26} \mathrm{cm^{-2}}$, decreasing to $\Sigma {\sim} 10^{23} \mathrm{cm^{-2}}$ at cylindrical radii of $400$~AU.
We reduce $\Sigma(y,z)$ to a 1D function by computing its azimuthal median, taking one sink's location as the coordinates origin; we obtain $\Sigma_\mathrm{median}(r_\mathrm{cyl})$.
We use a threshold criterion to identify the disk radius as the cylindrical radius within which the surface density $\Sigma_\mathrm{median}(r_\mathrm{cyl})$ exceeds a critical value taken to be $10^{25}\mathrm{cm^{-2}}$.
Using this diagnostic, we find that the disk grows in size, reaching $100$ to $150$~AU size in hydrodynamical runs before secondary fragmentation.
After secondary fragmentation, in which the newly-formed companion starts forming its own disk, the primary disk radius decreases and remains between $60$~AU and $100$~AU in runs {\tt HYDRO-HR} and {\tt HYDRO-VHR}.
During this epoch, the companion gains mass rapidly and the disk radius is compatible with disk truncation, e.g. about $0.3$ of the orbital separation (\citealt{papaloizou_tidal_1977}, \citealt{paczynski_model_1977}).
Meanwhile, in the magnetic runs, the disk radius remains generally smaller than $100$~AU even though no companion forms, as shown in Fig.~\ref{fig:rdisk}, and is compatible with magnetic regulation \citep{hennebelle_magnetically_2016}.
Varying the threshold value to be $5\times 10^{25} \mathrm{cm^{-2}}$ decreases the obtained disk radius by about $20$~AU.

Finally, \cite{commercon_discs_2022} found that the protostellar disks formed in presence of ambipolar diffusion have $\beta = P/P_\mathrm{mag} > 1$, where $P_\mathrm{mag}=B^2/(8 \pi)$ is the magnetic pressure - more recently, it has been shown in simulation incorporating Ohmic dissipation instead of ambipolar diffusion as well \citep{oliva_modeling_2023-1}.
 Following the former study, \cite{mignon-risse_collapse_2021} argued that this could serve as a way to measure the disk radius, especially when the dynamics and multiple stellar components make difficult the comparisons between rotation profiles and Keplerian ones, which is the case here. 
Here, we indeed found that the closest region to the sinks have a $\beta>1$ and that the transition with $\beta<1$ is located near the surface density transition setting the disk edge.
With a spatial resolution smaller than $5$~AU we derive a disk radius, based on the plasma beta, of $50$~AU to $80$~AU (see Fig.~\ref{fig:rdisk}).
This disk radius value is comparable to the surface density estimation using the largest surface density threshold among the two values explored here, and is nearly constant when varying the resolution.

To conclude, we found the column density map and the plasma beta to give a relatively compatible ($\pm20$~AU depending on the surface density) and converged (for resolution finer than $5$~AU) estimate of the disk radius.
Disks are roughly twice smaller in the magnetized case compared to the hydrodynamical case before secondary fragmentation.
They are compatible with magnetic regulation, while in the hydrodynamical case they grow until fragmentation occurs and the newly-formed companion truncates the primary disk.
Hence, in the hydrodynamical case, this result could be interpreted as the disk size not being regulated (it gathers material reaching its centrifugal limit) or being regulated by fragmentation.

%

\section{Discussion}

\subsection{Comparison with previous works}
\label{sec:comp_work}

{In the simulations presented here, the refinement criterion for the AMR levels below the maximum is a Jeans length resolved by $12$ cells.
While in the original work of \cite{truelove_jeans_1997}, a number of $4$ cells was mentioned (and widely used since then, see e.g. \citealt{kuruwita_binary_2017}, \citealt{gerrard_role_2019}; other authors opted for $8$ cells, see e.g. \citealt{masson_ambipolar_2016}, \citealt{rosen_unstable_2016}).
To achieve convergence with SPH methods, \cite{commercon_protostellar_2008} showed that in the hydrodynamical case the Jeans length should be sampled by at least $15$ cells.
In order to resolve minimal dynamo amplification in self-gravity MHD simulations, \cite{federrath_new_2011} showed that $30$ cells per Jeans length were required.
Because the present initial conditions are not turbulent and the present disks do not develop turbulent states, we chose to keep the Jeans length resolution to 12 cells to save computational time.}

{\cite{bate_diversity_2018} studied the morphology of protostellar disks from molecular clouds simulations with SPH methods, neglecting magnetic fields.
In our high-resolution hydrodynamical simulations (e.g. \textsc{HYDRO-VHR}), we obtain a quadruple system that is somehow similar to that depicted in their Fig.~4 (sinks 41, 89, 76, 83; see also \citealt{sigalotti_large-scale_2018}).
They have in common the following properties: this is a point-symmetric system composed of two tight pairs surrounded by gas.
Additionally, the two pairs are linked together by gas filaments continuously fed by the spiral arms belonging to protostellar disks.
Nevertheless, they report circumbinary disks around the tight pairs while we do not.
Instead, we find each star in the quadruple system to have a protostellar disk around it.
This could be a matter of angular momentum budget of the surrounding gas, since high angular momentum material is more prone to form circumbinary disks \citep{bate_accretion_1997}.
Finally, in our simulations, the youngest sinks formed from the interaction between the spiral arm belonging to the oldest sinks' disks and the filament linking them; such fragmentation triggered by arm-filament collision seems to occur in their study as well.
We find, however, that the inclusion of magnetic effects suppresses this fragmentation mode as magnetic pressure reduces the density contrast around the filament.
}

In this study, non-ideal MHD was included in the form of ambipolar diffusion.
Hence, let us compare the properties of the multiple stellar systems formed in the magnetized runs with those of the binary systems reported in \cite{mignon-risse_collapse_2021}, for which the included physics was identical.
The initial conditions are quantitatively different (core mass, rotation...) but the qualitative difference comes from the inclusion of initial velocity dispersion to mimic turbulence while here all the angular momentum originates from differential rotation.
Unlike \cite{mignon-risse_collapse_2021}, where a circumbinary structure is present under particular turbulence level and magnetic field strength, we find no circumbinary structure in the binary systems formed here, except in run {\tt MU2-VHR} where a sink is still orbiting within the accretion disk it was born in.
As in \cite{mignon-risse_collapse_2021}, we do find mass ratios of the order of unity in binary systems. 
This fits the "high angular momentum core" case, while a "low angular momentum core" is predicted to result in unequal-mass binaries by \cite{bate_accretion_1997-1}, \cite{bate_accretion_1997}.
The triple system in run {\tt MU2-VHR} has masses $3.7\, \mathrm{M_\odot}$, $3\, \mathrm{M_\odot}$ and $1.4\, \mathrm{M_\odot}$ at the end of the run, with a mass ratio close to one before the third sink formed.

Preliminary works in low-mass star formation with ambipolar diffusion reported disk sizes reduced with respect to the standard hydrodynamical case (\citealt{masson_ambipolar_2016}, \citealt{hennebelle_magnetically_2016}).
As in the studies of massive isolated collapse of \cite{commercon_discs_2022}, which forms a single-star system, and \cite{mignon-risse_collapse_2021}, which forms a binary system when initial turbulence dominates over magnetic fields, disks are found to be smaller in the non-ideal MHD case than in the hydrodynamical case, based on the column density maps (and on the plasma beta in the magnetic case).
Recently, \cite{lebreuilly_protoplanetary_2021} confirmed this trend on large samples in the context of collapses within massive ($1000\, \mathrm{M_\odot}$) clumps, aiming at describing the population of protoplanetary disks, although this work was mainly oriented towards low-mass star formation.

{\cite{gerrard_role_2019} studied the influence of an initialized turbulent magnetic field on disk fragmentation and outflow launching during the collapse of a low-mass pre-stellar core, under the ideal MHD approximation.
They claim that a turbulent magnetic field leads to a more isotropic magnetic pressure distribution than a non-turbulent one, ending up in a reduced magnetic pressure gradient and more disk fragmentation.
In fact, it can be expected from their assumptions (ideal MHD) that their disks are supported by magnetic pressure; while the disks formed here, in the presence of ambipolar diffusion, are thermally-supported, so that changes in the magnetic pressure distribution should not have a major influence on the disk fragmentation.
}

Magnetic fields may also play a role in the orbital separation of binaries.
Recently, \cite{harada_impact_2021} reported a numerical-and-analytical study of the impact of magnetic braking on massive close binaries, in the ideal MHD framework.
They show that the orbital separation is larger in the hydrodynamical case than when magnetic fields are introduced (their Eq.~6 with the non-specific angular momentum is equivalent to Eq.~\ref{eq:orbsep} here), because of magnetic braking and outflows.
Our findings qualitatively agree with this result (see Fig.~\ref{fig:t_b}), as we find an orbital separation reduced by a factor of ${\sim}2$ when magnetic fields are present.
However, we attribute it to the initial separation being smaller, because of early magnetic braking (when both stars were part of a coherent structure), while the further evolution of $\dot{a}/a$ is comparable; on the opposite, their results only apply to the urther evolution of $\dot{a}/a$, since the initial separation is fixed.
Moreover, quantitatively, we also find orbital separations larger than $100$~AU in the magnetic case, while they do not.
Our results differ because they assume a circumbinary structure able to extract angular momentum from the close binary, while we witness the formation of two separate disks except in the secondary-fragmentation formed binary system of run {\tt MU2-VHR}.
Such differences are partially attributed to the different choices of initial mass and angular momentum budget, which, as we have shown, drive the evolution of the orbital separation in binaries formed from initial fragmentation.
Our initial conditions are indeed different (magnetic field strength, mass and angular momentum budget). The MHD version is different as well: non-ideal here, ideal in their work.
Ideal MHD is found to overestimate the magnetic field strength in the densest ($\rho\gtrsim 10^{-15} \mathrm{g \, cm^{-3}}$) regions \citep{masson_ambipolar_2016}, impacting both disk formation and outflows \citep{commercon_discs_2022}.
Nevertheless, the present study supports the interpretation of \cite{harada_impact_2021} that magnetic fields, in the form of magnetic braking, could play a prominent role in the formation of massive close binaries -- at the condition of a coherent rotation structure such as a circumbinary disk --, at least initially.
Further in time, however, gas accretion equally widens binaries in the hydrodynamical and in the magnetized cases, in the absence of a circumbinary accretion structure.

{The formation of a circumbinary disk structure around a low-mass protobinary system in non-turbulent and turbulent cores has been studied by  \cite{kuruwita_role_2019}, including ideal MHD.
They report orbital shrinkage and attribute it to magnetic braking on the infalling gas.
Alternatively, this could be due to the sinks not being at centrifugal equilibrium when they form, since they arise from initial perturbations, shrinking their separation until they reach a stable orbit.
To understand whether the final evolution of their orbital separation is consistent with gas accretion (that is, no outspiral), we put into Eq.~\ref{eq:jgas} the final mass (${\sim}0.5\, \mathrm{M_\odot}$) and orbital separation (${\sim}20$~AU) of the protobinary system in their non-turbulent run.
This gives a minimum gas angular momentum of $2\times 10^{20} \, \mathrm{cm^2 \, s^{-1}}$ to drive the binary outspiral through gas accretion; meanwhile, their protostellar core has a maximal specific angular momentum of ${\approx}10^{18} \,\mathrm{cm^2  \, s^{-1}}$ at the core edge.
Hence, their pre-stellar core angular momentum budget is insufficient to drive the binary's outspiral.
Because they consider relatively small turbulent velocities ($0.02 \, \mathrm{km \, s^{-1} }$ and $0.04 \, \mathrm{km \, s^{-1} }$), the angular momentum carried out by turbulence is still insufficient to drive the binary's outspiral and this naturally results in a similar final orbital separation in their non-turbulent and turbulent runs.
Finally, in contrast to their turbulent runs, we do not report any circumbinary disk formation other than the one in run {\tt MU2-VHR} where a companion has previously formed. This could be due to the large ($>100$~AU) orbital separation in our runs compared to theirs (${\sim}20$~AU).
}

We have investigated the impact of multiplicity on the accretion rate, whose understanding is missing in the star formation context. 
Indeed, it is not a free parameter but rather an outcome of the self-consistent collapse, even in the studies dedicated to multiple system formation (e.g. \citealt{harada_impact_2021}). Nonetheless, it has been investigated for binary compact objects (see e.g. \citealt{bowen_quasi-periodic_2018}), which are already introduced in the initial conditions. Periodicities in the accretion rate, such as those we investigate in Sec.~\ref{sec:B_conv}, have been reported by \cite{bowen_quasi-periodic_2018}, but those are dominated by the periodic inflow from circumbinary disks.
Our present work suggests that no circumbinary disk is needed to modulate the accretion rate at orbital-like frequencies.
{This is in line with the episodes of enhanced accretion at periastron passage reported in \cite{kuruwita_role_2019}.}

Finally, let us mention the differences with Paper I in terms of multiplicity and disk radius.
First, it was shown in Paper I that the multiplicity outcome of the system is sensitive to numerical choices such as the sink properties (number, possibility to merge) and numerical grid as it can seed instabilities leading to fragmentation.
Here we have chosen a numerical setup favoring multiple system formation in order to study the influence of magnetic fields on its properties; the approach is somehow similar to \cite{bate_accretion_1997-1}, except that here the sinks formed self-consistently (within the uncertainty related to numerical choices as explicited above).
We have checked that the absence of an initial sink particle also leads to a binary system after a few kyr, on which our analysis is performed, and to qualitatively-similar results, which strenghens the results obtained here.
This is due to the several mergers presented in Sec.~\ref{sec:mult}, as mergers are triggered when the multiplicity is higher than $2$ in the innermost region of the cloud); for a higher multiplicity to be reached, hierarchical fragmentation is necessary in the present simulations.
This study also illustrates how fragmentation, even with an initial central sink particle, leads to symmetry breaking with respect to the axissymmetric initial conditions; allowing the central sink particle to move is key here not to enforce the central symmetry.
The present results on the orbital separation also bring additional clues to the debate regarding the link between the early phases and the multiplicity outcome.
Indeed, this link between early-phase, central fragmentation and the ultimate fate of the system is due to the orbital separation increasing, driving stars apart in the form of a multiple system.
We show here that such a trend depends on the initial angular momentum and mass distribution of the cloud because the outspiral is driven by gas accretion.
Second, we notice that the single disk in Paper I is larger than the disks formed in the binary systems of the {\tt HYDRO-} runs.
In those runs exhibiting a binary system (and in the magnetic runs as well), the disk radius is smaller than the tidal truncation radius, which is ${\sim}0.3$ times the orbital separation for equal-mass binaries (\citealt{papaloizou_tidal_1977}, \citealt{paczynski_model_1977}).
Hence, the disk radius is not attributable to tidal truncation by the companion before the secondary fragmentation epoch.
This is most likely due to hierarchical fragmentation, in which the specific angular momentum of the fragments is smaller than that of the parent cloud. 
Hierarchical fragmentation was already mentioned as a possibility to solve the angular momentum problem in star formation (see e.g. \citealt{bodenheimer_evolution_1978}, \citealt{zinnecker_star_1984}).
After secondary fragmentation however, as the multiplicity becomes larger than $2$, the disks are most likely truncated by tidal forces as they extend from one sink to the companion's disk (particularly visible for the {\tt HYDRO-HR} and {\tt HYDRO-VHR} runs in Fig.~\ref{fig:coldens}).

\subsection{Comparison with observations}

The initial specific angular momentum profile $j\varpropto R^{1.25}$ (see Sec.~\ref{sec:frag_ci}), where we recall that $R$ is the cylindrical radius, is motivated by theoretical purposes (\citealt{meyer_forming_2018}, \citealt{meyer_episodic_2019}).
It may not be representative of typical cores - if such massive pre-stellar cores do exist (see e.g. \citealt{zhang_probing_2020} for candidates and \citealt{tan_massive_2014}, \citealt{motte_high-mass_2018}, for recent reviews).
For instance, \citealt{pineda_specific_2019} measured $j\varpropto r^{1.8}$ in low-mass dense cores, which is not so different from the measurements of \cite{goodman_dense_1993} who found a specific total angular momentum $J/M \varpropto r^{1.6}$ (the slope of $j$ and $J/M$ are the same for monotonic profiles, see \citealt{pineda_specific_2019}).
Meanwhile, it could be important in setting the angular momentum versus mass budget and therefore the evolution of the orbital separation of binaries, as studied in Sec.~\ref{sec:orbsep}.
Nevertheless, this reasoning relies on a high-mass star formation process similar to that of low-mass stars. Large-scale dynamics, such as hierarchical collapse, should be taken into account, but this is beyond the scope of this paper.

We find that, with and without magnetic fields, primary disks form, fragment, and those fragments eventually possess their own disk as well. This hierarchical picture for fragmentation is supported by the recent observations of \cite{suri_disk_2021} in AFGL 2591.
Furthermore, the multiplicity of outflows in AFGL 2591 has been used by \cite{gieser_chemical_2019} to infer the presence of fragmented cores. 
Even though outflows are not the main topic of the present paper, let us note that we find magnetic outflows here as in \cite{mignon-risse_collapse_2021-1}, launched from individual objects in multiple systems (here represented as sink particles) while the collapse is still on-going. 
Therefore, it confirms magnetic outflows (namely, magnetic tower flows and magneto-centrifugal outflows, as studied in \citealt{mignon-risse_collapse_2021-1}), as relevant candidates for the protostellar outflows observed in multiple protostellar systems.
The comparison of polarized emission maps produced in the outflow region with observations is beyond the scope of this paper.

As already reported in \cite{mignon-risse_collapse_2021}, we find a prominent role of spiral arms in the formation of (massive) multiple stellar systems. Interestingly, \cite{tobin_triple_2016} found evidence that the triple protostellar system L1448 IRS3B is still embedded in the spiral arms it may have formed from. This is consistent with the present picture, if low- and high-mass multiple stellar systems share similarities in their accretion process.

Let us note that we do not report runaway stars in those simulations.
In runs {\tt HYDRO-} (of all resolutions) and {\tt MU2-VHR}, the multiplicity is strictly larger than two, so $N-$body interactions could possibly eject stars if the computation was integrated for a longer time.
Further work is needed to reconcile massive star formation simulations with the high frequency of massive runaway stars reported by observations (see e.g. \citealt{dorigo_jones_runaway_2020} and references therein).

A large fraction of massive stars are in close binaries (e.g. \citealt{sana_binary_2012}). 
While dynamical interactions have been shown to be a relevant process to form close (${<}10$~AU) binaries \citep{bate_formation_2002}, our run {\tt MU2-VHR} shows that disk fragmentation of magnetized disks could be a possible pathway. Indeed, we find an initial orbital separation of $19$~AU (which remained nearly constant for the short integrated time, namely $16$~AU after $1.7$~kyr) and a common accretion structure.
For a possible orbital shrinking to be followed, we would need finer resolution (here, $5$~AU is the sink radius).
We leave this to further work.

On the opposite, an excess of "twin" (mass ratio between $0.95$ and $1$) low-mass binaries with orbital separations ${>}1000$~AU have been reported in \cite{el-badry_discovery_2019} using the Gaia DR2 catalog.
This population is unlikely to arise from close binaries widened by positive torques from a circumbinary disk, since no circumbinary disk that large has ever been observed so far, in agreement with the non-magnetized large-scale numerical simulations of \cite{bate_diversity_2018}.
While our results should be scaled-down to low-mass binaries with care, they support gas accretion as a plausible candidate to help forming such a population, after an initial fragmentation phase which could occur within the innermost region of the pre-stellar core.
Indeed, we find gas accretion to widen binaries from a few tens AU to $500$~AU in the magnetized case with no sign of decrease and with an explicit criterion for further widening explicited in Eq.\ref{eq:jgas}.

\section{Conclusions}

In this paper, we have investigated the influence of spatial resolution and of magnetic fields on disk fragmentation and on the subsequent formation of massive multiple stellar systems.

We considered eight runs with no restriction on the number of sink particles to be formed, and one sink already present at the center initially. We studied the properties of the multiple stellar systems, varying the resolution and the presence of magnetic fields.
At early stages, a density bump-like structure forms after the adiabatic stage and fragments into several sink particles, breaking the axisymmetry via Toomre instability.
The sink initially present interacts gravitationally with the other sinks and leaves the center of the cloud.
When more than two sinks are present, their gravitational interactions induce radial migration with respect to the center of the cloud.
The sinks gaining (losing) angular momentum migrate to larger (smaller) distance from the center, until mergers occur with the other sinks and the multiplicity reduces to $2$, in all runs.
The orbital separation in binaries increases as they accrete gas possessing angular momentum with respect to the center of mass of the binary, e.g. the center of the cloud. 
The separation is decreased by a factor of two when magnetic fields are included, as an heritage of their initial orbital separation reduced via magnetic braking compared to the hydrodynamical case.
Hence, magnetic fields are promising to explain the formation of close binaries{, if there is a common accretion structure}.

The increasing orbital separation is driven by accretion of sufficiently high angular momentum gas.
Such a trend in the orbital separation is likely to depend on the initial rotation profile.
For initial conditions promoting gas-driven outspiral, any fragmentation occuring in the early phases is crucial for the final fate of the system, as briefly presented in Paper I.

A second generation of stars forms from disk fragmentation around the first generation of stars. 
In more details, fragmentation can be triggered via arm-arm collision or arm-filament collision; in the magnetic case, arm-filament collision is absent.
Our results are consistent with a hierarchical picture of fragmentation, in the hydrodynamical case and in the magnetic case (as in, e.g., \citealt{park_population_2022}).
We have shown that a resolution of $5$~AU gives a converged picture in the hydrodynamical case and $2.5$~AU in the magnetic case, i.e. about $1/20$ of the disk size, in order to resolve structures prone to fragmentation (filaments and spiral arms).
Those values depend on the size of the first generation's disks, which are found here to be about twice smaller in the magnetic runs as compared to the hydrodynamical runs, before secondary fragmentation.
We only observe secondary fragmentation in the magnetic case for a resolution of $1.25$~AU, suggesting that a lack of fragmentation could be, in this particular case, a lack of resolution.
Those results will serve as references for future numerical studies.
On the opposite, the disk sizes we find here recommend that any resolution above ${\sim}100$~AU simply misses disk fragmentation.

Even at low resolution ($10$~AU here), the convergence regarding the total sink mass ($15\%$) and the orbital separation of binary systems is correct.
In the accretion rates of binaries we find hints of modulations at the orbital frequency. 
This suggests that their tidal forces help the accretion onto the other object.
If confirmed by further work, it means that fragmentation does not limit the mass of stars but rather enhances accretion bursts and possibly increases the efficiency of gas conversion into stars in multiple systems.
This is of particular interest, since massive stars are more often in multiple systems than low-mass stars (e.g. \citealt{chini_spectroscopic_2012}).
This moderates the argument of the fragmentation-induced starvation scenario \citep{peters_limiting_2010}, in which fragmentation reduces the maximal stellar mass by sub-dividing the mass reservoir into several lower-mass stars.
\newline

Overall, this work suggests that the influence of magnetic fields in the formation of multiple masse stars - without even focusing on the outflows, which are now understood as mainly magnetic - is important in several aspects.
We find that magnetic fields regulate the disk size, which appears to be, otherwise, only regulated through disk fragmentation producing a companion, and tidal truncation by the same companion.
Among the possible fragmentation modes (arm-arm collision, arm-filament collision), we find the arm-filament collision mode to be suppressed because the filaments are more diffuse in the presence of magnetic fields.
Finally, magnetic braking reduces the orbital separation of binaries by a factor of two because i) it reduces their initial separation and ii) it has no impact on the further evolution of the orbital separation, driven through accretion.
In the absence of a common accretion structure, magnetic effects are insufficient to form close binaries.
For their strong impact on the disk growth and fragmentation and on the initial orbital separation, we conclude that magnetic fields are important in the formation of multiple massive stars.

\begin{acknowledgements}
	RMR thanks A. Oliva and R. Kuiper for useful discussion.
      This work was supported by the CNRS "Programme National de Physique Stellaire" (\emph{PNPS}). 
      This work has received funding from the French Agence Nationale de la Recherche (ANR) through the project COSMHIC (ANR-20-CE31-0009).
      The numerical simulations we have presented in this paper were produced on the \emph{CEA} machine \emph{Alfv\'en} and using HPC resources from GENCI-CINES (Grant A0080407247). The visualisation of \ramses{} data has been done with the \href{https://github.com/nvaytet/osyris}{OSYRIS} python package, for which RMR thanks Neil Vaytet.
\end{acknowledgements}

\begin{appendix}

\section{Computational cost and carbon footprint estimate}
\label{app:cost}

Table~\ref{table:cost} gives the computational cost of the simulations presented in Table~\ref{table:ics}.
One can notice the higher cost of MHD simulations. This is mainly due to the ambipolar diffusion timestep which is prohibitive for long-integration runs.
As the AMR grid refines regions of interest, in particular around stellar companions, the cost does not strictly scale with the finest resolution, as could be expected.
Simulations have been performed over $64$ CPU cores, except run {\tt MU2-VHR}, on $128$ CPU cores.

The $\mathrm{CO_{2,e}}$ ($\mathrm{CO_{2}}$ equivalent) carbon footprint has been computed using the estimate of $4.68\, \mathrm{g/CPUh}$ \citep{berthoud_estimation_2020}.

\begin{table}
\caption{Computational cost (in CPUkhr) and $\mathrm{CO_{2,e}}$ footprint estimate (in kg) of the simulations presented in Table~\ref{table:ics}.}
\label{table:cost}
\centering 
\begin{tabular}{c | c c} 
	\hline\hline
	 Model 				& Cost [CPUkhr] 	&   $\mathrm{CO_2}$ emission [kg] \\ \hline 
	{\tt HYDRO-LR}   	&  $1$	 & $4.68$	\\ 
	{\tt HYDRO-MR}		 &  $3$   	 &  $14.04$ 	  \\ 
	{\tt HYDRO-HR}		 &  $6$   &  $28.08$	   \\ 
	{\tt HYDRO-VHR}	 &  $11$   &  $51.48$	  \\  \hline
	{\tt MU2-LR}    		&   $8$   &    $37.44$	  \\ 
	{\tt MU2-MR}   		 &  $48$  &     $224.64$	   \\
	{\tt MU2-HR}   		 &  $78$   &    $365.04$	 \\ 
	{\tt MU2-VHR}   		 &  $125$   &    $585.00$  \\ \hline
	Total					& $280$ 	& $1310.4$ \\ \hline\hline
\end{tabular}
\end{table}

\end{appendix}

\bibliographystyle{aa} 
\bibliography{Zotero} 

\begin{thebibliography}{107}
\expandafter\ifx\csname natexlab\endcsname\relax\def\natexlab#1{#1}\fi

\bibitem[{Adams {et~al.}(1989)Adams, Ruden, \& Shu}]{adams_eccentric_1989}
Adams, F.~C., Ruden, S.~P., \& Shu, F.~H. 1989, ApJ, 347, 959

\bibitem[{Ali \& Harries(2019)}]{ali_massive_2019}
Ali, A.~A. \& Harries, T.~J. 2019, Monthly Notices of the Royal Astronomical
  Society, 487, 4890, arXiv: 1906.05858

\bibitem[{Alves {et~al.}(2017)Alves, Girart, Caselli, Franco, Zhao, Vlemmings,
  Evans, \& Ricci}]{alves_molecular_2017}
Alves, F.~O., Girart, J.~M., Caselli, P., {et~al.} 2017, A\&A, 603, L3

\bibitem[{Alves {et~al.}(2018)Alves, Girart, Padovani, Galli, Franco, Caselli,
  Vlemmings, Zhang, \& Wiesemeyer}]{alves_magnetic_2018}
Alves, F.~O., Girart, J.~M., Padovani, M., {et~al.} 2018, A\&A, 616, A56

\bibitem[{Añez-López {et~al.}(2020)Añez-López, Busquet, Koch, Girart, Liu,
  Santos, Chapman, Novak, Palau, Ho, \& Zhang}]{anez-lopez_role_2020}
Añez-López, N., Busquet, G., Koch, P.~M., {et~al.} 2020, arXiv:2010.13503
  [astro-ph], arXiv: 2010.13503

\bibitem[{Bate(1997)}]{bate_accretion_1997-1}
Bate, M.~R. 1997, MNRAS, 285, 16

\bibitem[{Bate(2000)}]{bate_predicting_2000}
Bate, M.~R. 2000, Monthly Notices of the Royal Astronomical Society, 314, 33

\bibitem[{Bate(2018)}]{bate_diversity_2018}
Bate, M.~R. 2018, Monthly Notices of the Royal Astronomical Society, 475, 5618

\bibitem[{Bate \& Bonnell(1997)}]{bate_accretion_1997}
Bate, M.~R. \& Bonnell, I.~A. 1997, Monthly Notices of the Royal Astronomical
  Society, 285, 33

\bibitem[{Bate {et~al.}(2002)Bate, Bonnell, \& Bromm}]{bate_formation_2002}
Bate, M.~R., Bonnell, I.~A., \& Bromm, V. 2002, MNRAS, 336, 705

\bibitem[{Berthoud {et~al.}(2020)Berthoud, Bzeznik, Gibelin, Laurens, Bonamy,
  Morel, \& Schwindenhammer}]{berthoud_estimation_2020}
Berthoud, F., Bzeznik, B., Gibelin, N., {et~al.} 2020, Estimation de
  l'empreinte carbone d'une heure.coeur de calcul

\bibitem[{Beuther {et~al.}(2020)Beuther, Soler, Linz, Henning, Gieser, Kuiper,
  Vlemmings, Hennebelle, Feng, Smith, \& Ahmadi}]{beuther_gravity_2020}
Beuther, H., Soler, J.~D., Linz, H., {et~al.} 2020, arXiv:2010.05825
  [astro-ph], arXiv: 2010.05825

\bibitem[{Bleuler \& Teyssier(2014)}]{bleuler_towards_2014}
Bleuler, A. \& Teyssier, R. 2014, MNRAS, 445, 4015

\bibitem[{Bodenheimer(1978)}]{bodenheimer_evolution_1978}
Bodenheimer, P. 1978, ApJ, 224, 488

\bibitem[{Bonnell \& Bate(1994)}]{bonnell_massive_1994}
Bonnell, I.~A. \& Bate, M.~R. 1994, MNRAS, 269, L45

\bibitem[{Bowen {et~al.}(2018)Bowen, Mewes, Campanelli, Noble, Krolik, \&
  Zilhão}]{bowen_quasi-periodic_2018}
Bowen, D.~B., Mewes, V., Campanelli, M., {et~al.} 2018, ApJ, 853, L17

\bibitem[{Chini {et~al.}(2012)Chini, Hoffmeister, Nasseri, Stahl, \&
  Zinnecker}]{chini_spectroscopic_2012}
Chini, R., Hoffmeister, V.~H., Nasseri, A., Stahl, O., \& Zinnecker, H. 2012,
  Monthly Notices of the Royal Astronomical Society, 424, 1925

\bibitem[{Commerçon {et~al.}(2014)Commerçon, Debout, \&
  Teyssier}]{commercon_fast_2014}
Commerçon, B., Debout, V., \& Teyssier, R. 2014, A\&A, 563, A11

\bibitem[{Commerçon {et~al.}(2022)Commerçon, González, Mignon-Risse,
  Hennebelle, \& Vaytet}]{commercon_discs_2022}
Commerçon, B., González, M., Mignon-Risse, R., Hennebelle, P., \& Vaytet, N.
  2022, A\&A, 658, A52

\bibitem[{Commerçon {et~al.}(2008)Commerçon, Hennebelle, Audit, Chabrier, \&
  Teyssier}]{commercon_protostellar_2008}
Commerçon, B., Hennebelle, P., Audit, E., Chabrier, G., \& Teyssier, R. 2008,
  A\&A, 482, 371

\bibitem[{Commerçon {et~al.}(2010)Commerçon, Hennebelle, Audit, Chabrier, \&
  Teyssier}]{commercon_protostellar_2010}
Commerçon, B., Hennebelle, P., Audit, E., Chabrier, G., \& Teyssier, R. 2010,
  A\&A, 510, L3

\bibitem[{Commerçon {et~al.}(2011{\natexlab{a}})Commerçon, Hennebelle, \&
  Henning}]{commercon_collapse_2011}
Commerçon, B., Hennebelle, P., \& Henning, T. 2011{\natexlab{a}}, ApJ, 742, L9

\bibitem[{Commerçon {et~al.}(2011{\natexlab{b}})Commerçon, Teyssier, Audit,
  Hennebelle, \& Chabrier}]{commercon_radiation_2011}
Commerçon, B., Teyssier, R., Audit, E., Hennebelle, P., \& Chabrier, G.
  2011{\natexlab{b}}, A\&A, 529, A35

\bibitem[{Cox {et~al.}(2022)Cox, Novak, Sadavoy, Looney, Lee, Berthoud, Bourke,
  Coudé, Encalada, Fissel, Harrison, Houde, Li, Myers, Pattle, Santos,
  Stephens, Wang, \& Wolf}]{cox_twisted_2022}
Cox, E.~G., Novak, G., Sadavoy, S., {et~al.} 2022, The {Twisted} {Magnetic}
  {Field} of the {Protobinary} {L483}, arXiv:2206.00683 [astro-ph]

\bibitem[{Dorigo~Jones {et~al.}(2020)Dorigo~Jones, Oey, Paggeot, Castro, \&
  Moe}]{dorigo_jones_runaway_2020}
Dorigo~Jones, J., Oey, M.~S., Paggeot, K., Castro, N., \& Moe, M. 2020, ApJ,
  903, 43

\bibitem[{Duchêne \& Kraus(2013)}]{duchene_stellar_2013}
Duchêne, G. \& Kraus, A. 2013, Annu. Rev. Astron. Astrophys., 51, 269

\bibitem[{El-Badry {et~al.}(2019)El-Badry, Rix, Tian, Duchêne, \&
  Moe}]{el-badry_discovery_2019}
El-Badry, K., Rix, H.-W., Tian, H., Duchêne, G., \& Moe, M. 2019, Monthly
  Notices of the Royal Astronomical Society, 489, 5822, arXiv:1906.10128
  [astro-ph]

\bibitem[{Federrath \& Klessen(2012)}]{federrath_star_2012}
Federrath, C. \& Klessen, R.~S. 2012, ApJ, 761, 156

\bibitem[{Federrath {et~al.}(2011)Federrath, Sur, Schleicher, Banerjee, \&
  Klessen}]{federrath_new_2011}
Federrath, C., Sur, S., Schleicher, D. R.~G., Banerjee, R., \& Klessen, R.~S.
  2011, ApJ, 731, 62

\bibitem[{Fromang {et~al.}(2006)Fromang, Hennebelle, \&
  Teyssier}]{fromang_high_2006}
Fromang, S., Hennebelle, P., \& Teyssier, R. 2006, A\&A, 457, 371

\bibitem[{Galametz {et~al.}(2020)Galametz, Maury, Girart, Rao, Zhang, Gaudel,
  Valdivia, Hennebelle, Cabedo-Soto, Keto, \&
  Lai}]{galametz_observational_2020}
Galametz, M., Maury, A., Girart, J.~M., {et~al.} 2020, arXiv:2010.12466
  [astro-ph], arXiv: 2010.12466

\bibitem[{Galli \& Shu(1993)}]{galli_collapse_1993}
Galli, D. \& Shu, F.~H. 1993, ApJ, 417, 220

\bibitem[{Gerrard {et~al.}(2019)Gerrard, Federrath, \&
  Kuruwita}]{gerrard_role_2019}
Gerrard, I.~A., Federrath, C., \& Kuruwita, R. 2019, Monthly Notices of the
  Royal Astronomical Society, 485, 5532

\bibitem[{Gieser {et~al.}(2019)Gieser, Semenov, Beuther, Ahmadi, Mottram,
  Henning, Beltran, Maud, Bosco, Leurini, Peters, Klaassen, Kuiper, Feng,
  Urquhart, Moscadelli, Csengeri, Lumsden, Winters, Suri, Zhang, Pudritz,
  Palau, Menten, Galvan-Madrid, Wyrowski, Schilke, Sánchez-Monge, Linz,
  Johnston, Jiménez-Serra, Longmore, \& Möller}]{gieser_chemical_2019}
Gieser, C., Semenov, D., Beuther, H., {et~al.} 2019, A\&A, 631, A142

\bibitem[{Girart {et~al.}(2009)Girart, Beltran, Zhang, Rao, \&
  Estalella}]{girart_magnetic_2009}
Girart, J.~M., Beltran, M.~T., Zhang, Q., Rao, R., \& Estalella, R. 2009,
  Science, 324, 1408

\bibitem[{Girart {et~al.}(2013)Girart, Frau, Zhang, Koch, Qiu, Tang, Lai, \&
  Ho}]{girart_dr_2013}
Girart, J.~M., Frau, P., Zhang, Q., {et~al.} 2013, ApJ, 772, 69

\bibitem[{Goodman {et~al.}(1993)Goodman, Benson, Fuller, \&
  Myers}]{goodman_dense_1993}
Goodman, A.~A., Benson, P.~J., Fuller, G.~A., \& Myers, P.~C. 1993, The
  Astrophysical Journal, 406, 528

\bibitem[{Grudić \& Hopkins(2018)}]{grudic_elephant_2018}
Grudić, M.~Y. \& Hopkins, P.~F. 2018, arXiv:1809.08344 [astro-ph], arXiv:
  1809.08344

\bibitem[{Grudić {et~al.}(2020)Grudić, Kruijssen, Faucher-Giguère, Hopkins,
  Ma, Quataert, \& Boylan-Kolchin}]{grudic_model_2020}
Grudić, M.~Y., Kruijssen, J. M.~D., Faucher-Giguère, C.-A., {et~al.} 2020,
  arXiv:2008.04453 [astro-ph], arXiv: 2008.04453

\bibitem[{Gutiérrez-Vera {et~al.}(2022)Gutiérrez-Vera, Grassi, Bovino, Lupi,
  Galli, \& Schleicher}]{gutierrez-vera_non-ideal_2022}
Gutiérrez-Vera, N., Grassi, T., Bovino, S., {et~al.} 2022, Non-ideal
  magnetohydrodynamics of self-gravitating filaments, arXiv:2211.15615
  [astro-ph]

\bibitem[{Harada {et~al.}(2021)Harada, Hirano, Machida, \&
  Hosokawa}]{harada_impact_2021}
Harada, N., Hirano, S., Machida, M.~N., \& Hosokawa, T. 2021, Monthly Notices
  of the Royal Astronomical Society, 508, 3730

\bibitem[{Hennebelle {et~al.}(2016)Hennebelle, Commerçon, Chabrier, \&
  Marchand}]{hennebelle_magnetically_2016}
Hennebelle, P., Commerçon, B., Chabrier, G., \& Marchand, P. 2016, ApJ, 830,
  L8

\bibitem[{Hennebelle {et~al.}(2020)Hennebelle, Commerçon, Lee, \&
  Charnoz}]{hennebelle_what_2020}
Hennebelle, P., Commerçon, B., Lee, Y.-N., \& Charnoz, S. 2020, A\&A, 635, A67

\bibitem[{Hennebelle \& Teyssier(2008)}]{hennebelle_magnetic_2008-1}
Hennebelle, P. \& Teyssier, R. 2008, A\&A, 477, 25

\bibitem[{Ilee {et~al.}(2018)Ilee, Cyganowski, Brogan, Hunter, Forgan, Haworth,
  Clarke, \& Harries}]{ilee_g1192061_2018}
Ilee, J.~D., Cyganowski, C.~J., Brogan, C.~L., {et~al.} 2018, ApJ, 869, L24

\bibitem[{Jeans(1902)}]{jeans_vibrations_1902}
Jeans, J.~H. 1902, Proceedings of the Royal Society of London Series I, 71, 136

\bibitem[{Johnston {et~al.}(2020{\natexlab{a}})Johnston, Hoare, Beuther,
  Kuiper, Kee, Linz, Boley, Maud, Ahmadi, \& Robitaille}]{johnston_spiral_2020}
Johnston, K.~G., Hoare, M.~G., Beuther, H., {et~al.} 2020{\natexlab{a}}, A\&A,
  634, L11

\bibitem[{Johnston {et~al.}(2020{\natexlab{b}})Johnston, Hoare, Beuther, Linz,
  Boley, Kuiper, Kee, \& Robitaille}]{johnston_detailed_2020}
Johnston, K.~G., Hoare, M.~G., Beuther, H., {et~al.} 2020{\natexlab{b}},
  arXiv:2004.13739 [astro-ph], arXiv: 2004.13739

\bibitem[{Joos {et~al.}(2012)Joos, Hennebelle, \&
  Ciardi}]{joos_protostellar_2012}
Joos, M., Hennebelle, P., \& Ciardi, A. 2012, A\&A, 543, A128

\bibitem[{Kuruwita \& Federrath(2019)}]{kuruwita_role_2019}
Kuruwita, R.~L. \& Federrath, C. 2019, Monthly Notices of the Royal
  Astronomical Society, 486, 3647

\bibitem[{Kuruwita {et~al.}(2017)Kuruwita, Federrath, \&
  Ireland}]{kuruwita_binary_2017}
Kuruwita, R.~L., Federrath, C., \& Ireland, M. 2017, Monthly Notices of the
  Royal Astronomical Society, 470, 1626

\bibitem[{Kölligan \& Kuiper(2018)}]{kolligan_jets_2018}
Kölligan, A. \& Kuiper, R. 2018, A\&A, 620, A182

\bibitem[{Lebreuilly {et~al.}(2021)Lebreuilly, Hennebelle, Colman, Commerçon,
  Klessen, Maury, Molinari, \& Testi}]{lebreuilly_protoplanetary_2021}
Lebreuilly, U., Hennebelle, P., Colman, T., {et~al.} 2021, ApJL, 917, L10

\bibitem[{Levermore(1984)}]{levermore_relating_1984}
Levermore, C.~D. 1984, JQSRT, 31, 149

\bibitem[{Levermore \& Pomraning(1981)}]{levermore_flux-limited_1981}
Levermore, C.~D. \& Pomraning, G.~C. 1981, ApJ, 248, 321

\bibitem[{Machida {et~al.}(2005{\natexlab{a}})Machida, Matsumoto, Hanawa, \&
  Tomisaka}]{machida_collapse_2005}
Machida, M.~N., Matsumoto, T., Hanawa, T., \& Tomisaka, K. 2005{\natexlab{a}},
  MNRAS, 362, 382

\bibitem[{Machida {et~al.}(2005{\natexlab{b}})Machida, Matsumoto, Tomisaka, \&
  Hanawa}]{machida_collapse_2005-1}
Machida, M.~N., Matsumoto, T., Tomisaka, K., \& Hanawa, T. 2005{\natexlab{b}},
  MNRAS, 362, 369

\bibitem[{Marchand {et~al.}(2016)Marchand, Masson, Chabrier, Hennebelle,
  Commerçon, \& Vaytet}]{marchand_chemical_2016}
Marchand, P., Masson, J., Chabrier, G., {et~al.} 2016, Astronomy \&
  Astrophysics, 592, A18

\bibitem[{Masson {et~al.}(2016)Masson, Chabrier, Hennebelle, Vaytet, \&
  Commerçon}]{masson_ambipolar_2016}
Masson, J., Chabrier, G., Hennebelle, P., Vaytet, N., \& Commerçon, B. 2016,
  Astronomy \& Astrophysics, 587, A32

\bibitem[{Masson {et~al.}(2012)Masson, Teyssier, Mulet-Marquis, Hennebelle, \&
  Chabrier}]{masson_incorporating_2012}
Masson, J., Teyssier, R., Mulet-Marquis, C., Hennebelle, P., \& Chabrier, G.
  2012, The Astrophysical Journal Supplement Series, 201, 24

\bibitem[{Mathew \& Federrath(2021)}]{mathew_imf_2021}
Mathew, S.~S. \& Federrath, C. 2021, Monthly Notices of the Royal Astronomical
  Society, 507, 2448

\bibitem[{Meyer {et~al.}(2019)Meyer, Haemmerlé, \&
  Vorobyov}]{meyer_episodic_2019}
Meyer, D. M.~A., Haemmerlé, L., \& Vorobyov, E.~I. 2019, MNRAS, 484, 2482

\bibitem[{Meyer {et~al.}(2018)Meyer, Kuiper, Kley, Johnston, \&
  Vorobyov}]{meyer_forming_2018}
Meyer, D. M.-A., Kuiper, R., Kley, W., Johnston, K.~G., \& Vorobyov, E. 2018,
  MNRAS, 473, 3615

\bibitem[{Mignon-Risse {et~al.}(2021{\natexlab{a}})Mignon-Risse, González, \&
  Commerçon}]{mignon-risse_collapse_2021-1}
Mignon-Risse, R., González, M., \& Commerçon, B. 2021{\natexlab{a}}, A\&A,
  656, A85

\bibitem[{Mignon-Risse {et~al.}(2020)Mignon-Risse, González, Commerçon, \&
  Rosdahl}]{mignon-risse_new_2020}
Mignon-Risse, R., González, M., Commerçon, B., \& Rosdahl, J. 2020, A\&A,
  635, A42

\bibitem[{Mignon-Risse {et~al.}(2021{\natexlab{b}})Mignon-Risse, González,
  Commerçon, \& Rosdahl}]{mignon-risse_collapse_2021}
Mignon-Risse, R., González, M., Commerçon, B., \& Rosdahl, J.
  2021{\natexlab{b}}, A\&A, 652, A69

\bibitem[{Mignon-Risse {et~al.}(2023)Mignon-Risse, Oliva, González, Kuiper, \&
  Commerçon}]{mignon-risse_disk_2023}
Mignon-Risse, R., Oliva, A., González, M., Kuiper, R., \& Commerçon, B. 2023,
  publisher: arXiv Version Number: 1

\bibitem[{Miyoshi \& Kusano(2005)}]{miyoshi_multi-state_2005}
Miyoshi, T. \& Kusano, K. 2005, Journal of Computational Physics, 208, 315

\bibitem[{Moscadelli {et~al.}(2022)Moscadelli, Sanna, Beuther, Oliva, \&
  Kuiper}]{moscadelli_snapshot_2022}
Moscadelli, L., Sanna, A., Beuther, H., Oliva, A., \& Kuiper, R. 2022, Nat
  Astron, 6, 1068

\bibitem[{Motte {et~al.}(2018)Motte, Bontemps, \&
  Louvet}]{motte_high-mass_2018}
Motte, F., Bontemps, S., \& Louvet, F. 2018, Annu. Rev. Astron. Astrophys., 56,
  41

\bibitem[{Mouschovias \& Spitzer(1976)}]{mouschovias_note_1976}
Mouschovias, T.~C. \& Spitzer, Jr., L. 1976, The Astrophysical Journal, 210,
  326

\bibitem[{Murillo {et~al.}(2013)Murillo, Lai, Bruderer, Harsono, \& van
  Dishoeck}]{murillo_keplerian_2013}
Murillo, N.~M., Lai, S.-P., Bruderer, S., Harsono, D., \& van Dishoeck, E.~F.
  2013, A\&A, 560, A103

\bibitem[{Myers {et~al.}(2013)Myers, McKee, Cunningham, Klein, \&
  Krumholz}]{myers_fragmentation_2013}
Myers, A.~T., McKee, C.~F., Cunningham, A.~J., Klein, R.~I., \& Krumholz, M.~R.
  2013, The Astrophysical Journal, 766, 97

\bibitem[{Norman \& Wilson(1978)}]{norman_fragmentation_1978}
Norman, M.~L. \& Wilson, J.~R. 1978, ApJ, 224, 497

\bibitem[{Oliva \& Kuiper(2023{\natexlab{a}})}]{oliva_modeling_2023-1}
Oliva, A. \& Kuiper, R. 2023{\natexlab{a}}, A\&A, 669, A80

\bibitem[{Oliva \& Kuiper(2023{\natexlab{b}})}]{oliva_modeling_2023}
Oliva, A. \& Kuiper, R. 2023{\natexlab{b}}, A\&A, 669, A81

\bibitem[{Oliva \& Kuiper(2020)}]{oliva_modeling_2020}
Oliva, G.~A. \& Kuiper, R. 2020, A\&A, 644, A41

\bibitem[{Paczynski(1977)}]{paczynski_model_1977}
Paczynski, B. 1977, The Astrophysical Journal, 216, 822, aDS Bibcode:
  1977ApJ...216..822P

\bibitem[{Padoan {et~al.}(2014)Padoan, Federrath, Chabrier, Evans~II,
  Johnstone, Jørgensen, McKee, \& Nordlund}]{padoan_star_2014}
Padoan, P., Federrath, C., Chabrier, G., {et~al.} 2014, in Protostars and
  {Planets} {VI} (University of Arizona Press)

\bibitem[{Papaloizou \& Pringle(1977)}]{papaloizou_tidal_1977}
Papaloizou, J. \& Pringle, J.~E. 1977, Monthly Notices of the Royal
  Astronomical Society, 181, 441

\bibitem[{Park {et~al.}(2022)Park, Ricotti, \& Sugimura}]{park_population_2022}
Park, J., Ricotti, M., \& Sugimura, K. 2022, Population {III} star formation in
  an {X}-ray background: {III}. {Periodic} radiative feedback and luminosity
  induced by elliptical orbits, arXiv:2212.04564 [astro-ph]

\bibitem[{Patel {et~al.}(2005)Patel, Curiel, Sridharan, Zhang, Hunter, Ho,
  Torrelles, Moran, Gómez, \& Anglada}]{patel_disk_2005}
Patel, N.~A., Curiel, S., Sridharan, T.~K., {et~al.} 2005, Nature, 437, 109

\bibitem[{Peters {et~al.}(2010)Peters, Klessen, Low, \&
  Banerjee}]{peters_limiting_2010}
Peters, T., Klessen, R.~S., Low, M.-M.~M., \& Banerjee, R. 2010, The
  Astrophysical Journal, 725, 134

\bibitem[{Pineda {et~al.}(2019)Pineda, Zhao, Schmiedeke, Segura-Cox, Caselli,
  Myers, Tobin, \& Dunham}]{pineda_specific_2019}
Pineda, J.~E., Zhao, B., Schmiedeke, A., {et~al.} 2019, ApJ, 882, 103

\bibitem[{Rosdahl {et~al.}(2013)Rosdahl, Blaizot, Aubert, Stranex, \&
  Teyssier}]{rosdahl_ramses-rt:_2013}
Rosdahl, J., Blaizot, J., Aubert, D., Stranex, T., \& Teyssier, R. 2013, MNRAS,
  436, 2188

\bibitem[{Rosdahl \& Teyssier(2015)}]{rosdahl_scheme_2015}
Rosdahl, J. \& Teyssier, R. 2015, MNRAS, 449, 4380

\bibitem[{Rosen {et~al.}(2016)Rosen, Krumholz, McKee, \&
  Klein}]{rosen_unstable_2016}
Rosen, A.~L., Krumholz, M.~R., McKee, C.~F., \& Klein, R.~I. 2016, MNRAS, 463,
  2553

\bibitem[{Rosen {et~al.}(2021)Rosen, Offner, Foley, \&
  Lopez}]{rosen_blowing_2021}
Rosen, A.~L., Offner, S. S.~R., Foley, M.~M., \& Lopez, L.~A. 2021,
  arXiv:2107.12397 [astro-ph], arXiv: 2107.12397

\bibitem[{Sadavoy {et~al.}(2018)Sadavoy, Myers, Stephens, Tobin, Kwon,
  Segura-Cox, Henning, Commerçon, \& Looney}]{sadavoy_dust_2018}
Sadavoy, S.~I., Myers, P.~C., Stephens, I.~W., {et~al.} 2018, ApJ, 869, 115

\bibitem[{Sana {et~al.}(2012)Sana, de~Mink, de~Koter, Langer, Evans, Gieles,
  Gosset, Izzard, Le~Bouquin, \& Schneider}]{sana_binary_2012}
Sana, H., de~Mink, S.~E., de~Koter, A., {et~al.} 2012, Science, 337, 444

\bibitem[{Savonije {et~al.}(1994)Savonije, Papaloizou, \&
  Lin}]{savonije_tidally_1994}
Savonije, G.~J., Papaloizou, J. C.~B., \& Lin, D. N.~C. 1994, Monthly Notices
  of the Royal Astronomical Society, 268, 13

\bibitem[{Shu {et~al.}(1990)Shu, Tremaine, Adams, \& Ruden}]{shu_sling_1990}
Shu, F.~H., Tremaine, S., Adams, F.~C., \& Ruden, S.~P. 1990, ApJ, 358, 495

\bibitem[{Sigalotti {et~al.}(2018)Sigalotti, Cruz, Gabbasov, Klapp, \&
  Ramírez-Velasquez}]{sigalotti_large-scale_2018}
Sigalotti, L. D.~G., Cruz, F., Gabbasov, R., Klapp, J., \& Ramírez-Velasquez,
  J. 2018, ApJ, 857, 40

\bibitem[{Suri {et~al.}(2021)Suri, Beuther, Gieser, Ahmadi, Sánchez-Monge,
  Winters, Linz, Henning, Beltrán, Bosco, Cesaroni, Csengeri, Feng, Hoare,
  Johnston, Klaasen, Kuiper, Leurini, Longmore, Lumsden, Maud, Moscadelli,
  Möller, Palau, Peters, Pudritz, Ragan, Semenov, Schilke, Urquhart, Wyrowski,
  \& Zinnecker}]{suri_disk_2021}
Suri, S., Beuther, H., Gieser, C., {et~al.} 2021, arXiv:2109.04751 [astro-ph],
  arXiv: 2109.04751

\bibitem[{Tan {et~al.}(2014)Tan, Beltran, Caselli, Fontani, Fuente, Krumholz,
  McKee, \& Stolte}]{tan_massive_2014}
Tan, J.~C., Beltran, M.~T., Caselli, P., {et~al.} 2014, arXiv:1402.0919
  [astro-ph], arXiv: 1402.0919

\bibitem[{Teyssier(2002)}]{teyssier_cosmological_2002}
Teyssier, R. 2002, A\&A, 385, 337

\bibitem[{Tiede {et~al.}(2020)Tiede, Zrake, MacFadyen, \&
  Haiman}]{tiede_gas-driven_2020}
Tiede, C., Zrake, J., MacFadyen, A., \& Haiman, Z. 2020, ApJ, 900, 43

\bibitem[{Tobin {et~al.}(2016)Tobin, Kratter, Persson, Looney, Dunham,
  Segura-Cox, Li, Chandler, Sadavoy, Harris, Melis, \&
  Pérez}]{tobin_triple_2016}
Tobin, J.~J., Kratter, K.~M., Persson, M.~V., {et~al.} 2016, Nature, 538, 483

\bibitem[{Tobin {et~al.}(2020)Tobin, Sheehan, Reynolds, Megeath, Osorio,
  Anglada, Diaz-Rodriguez, Furlan, Kratter, Offner, Looney, Kama, Li, Hoff,
  Sadavoy, \& Karnath}]{tobin_vlaalma_2020}
Tobin, J.~J., Sheehan, P., Reynolds, N., {et~al.} 2020, arXiv:2011.01160
  [astro-ph], arXiv: 2011.01160

\bibitem[{Truelove {et~al.}(1997)Truelove, Klein, McKee, Holliman~II, Howell,
  \& Greenough}]{truelove_jeans_1997}
Truelove, J.~K., Klein, R.~I., McKee, C.~F., {et~al.} 1997, ApJ, 489, L179

\bibitem[{Tsukamoto {et~al.}(2022)Tsukamoto, Maury, Commerçon, Alves, Cox,
  Sakai, Ray, Zhao, \& Machida}]{tsukamoto_role_2022}
Tsukamoto, Y., Maury, A., Commerçon, B., {et~al.} 2022, publisher: arXiv
  Version Number: 1

\bibitem[{Wang {et~al.}(2019)Wang, Lai, Clemens, Koch, Eswaraiah, Chen, \&
  Pandey}]{wang_multiwavelength_2019}
Wang, J.-W., Lai, S.-P., Clemens, D.~P., {et~al.} 2019, arXiv:1911.11364
  [astro-ph], arXiv: 1911.11364

\bibitem[{Wurster \& Bate(2019)}]{wurster_disc_2019}
Wurster, J. \& Bate, M.~R. 2019, MNRAS

\bibitem[{Wurster \& Li(2018)}]{wurster_role_2018}
Wurster, J. \& Li, Z.-Y. 2018, Frontiers in Astronomy and Space Sciences, 5

\bibitem[{Zhang {et~al.}(2020)Zhang, Li, Pillai, Csengeri, Wyrowski, Menten, \&
  Pestalozzi}]{zhang_probing_2020}
Zhang, C.-P., Li, G.-X., Pillai, T., {et~al.} 2020, A\&A, 638, A105

\bibitem[{Zhang {et~al.}(2014)Zhang, Qiu, Girart, Liu, Tang, Koch, Li, Keto,
  Ho, Rao, Lai, Ching, Frau, Chen, Li, Padovani, Bontemps, Csengeri, \&
  Juárez}]{zhang_magnetic_2014}
Zhang, Q., Qiu, K., Girart, J.~M., {et~al.} 2014, The Astrophysical Journal,
  792, 116

\bibitem[{Zinnecker(1984)}]{zinnecker_star_1984}
Zinnecker, H. 1984, Monthly Notices of the Royal Astronomical Society, 210, 43

\end{thebibliography}

\end{document}